\documentclass[lettersize,journal]{IEEEtran}
\usepackage{amsmath,amsfonts}
\usepackage{array}
\usepackage[caption=false,font=normalsize,labelfont=sf,textfont=sf]{subfig}
\usepackage{cite}
\usepackage{amsmath,amssymb,amsfonts}
\usepackage{graphicx}
\usepackage{textcomp}
\usepackage{xcolor}
\usepackage{algorithm}
\usepackage{stfloats}
\usepackage{booktabs}
\usepackage{ulem}
\usepackage{threeparttable}
\usepackage{algorithm}
\usepackage{algpseudocode}
\usepackage{amsmath, amssymb}
\algrenewcommand\algorithmicrequire{\textbf{Input:}}
\algrenewcommand\algorithmicensure{\textbf{Output:}}
\algnewcommand{\LineComment}[1]{\State \(\triangleright\) #1}
\usepackage{caption}
\usepackage[caption=false,font=normalsize,labelfont=sf,textfont=sf]{subfig}
\def\BibTeX{{\rm B\kern-.05em{\sc i\kern-.025em b}\kern-.08em
    T\kern-.1667em\lower.7ex\hbox{E}\kern-.125emX}}
\usepackage{balance}
\begin{document}
\title{{Multi-Agent-Driven Cognitive Secure Communications in Satellite-Terrestrial Networks}}
\author{Yujie Ling,~\IEEEmembership{Graduate Student Member,~IEEE}, Zan Li,~\IEEEmembership{Fellow,~IEEE}, Lei Guan, Zheng Zhang,\\
Shengyu Zhang,~\IEEEmembership{Member,~IEEE}, and Tony Q.S. Quek,~\IEEEmembership{Fellow,~IEEE} \vspace{-1.2em}  
\thanks{This work was supported by the National Natural Science Foundation of China 62425103, XPLORER PRIZE, Innovative Research Groups of the National Natural Science Foundation of China 62121001, Discipline Innovation and Talent Introduction Base of Colleges and Universities in Shaanxi Province. \textit{(Corresponding author: Zan Li.)}

Yujie Ling, Zan Li, Lei Guan, and Zheng Zhang are with the State Key Laboratory of Integrated Services Networks, Xidian University, Xi'an 710071, China (e-mail: yujieling@stu.xidian.edu.cn; zanli@xidian.edu.cn; lguan@xidian.edu.cn; zhangzheng@xidian.edu.cn).

Shengyu Zhang, and Tony Q.S. Quek are with the Singapore University of Technology and Design, Singapore 487372 (e-mail: shengyu\_zhang@sutd.edu.sg; tonyquek@sutd.edu.sg).

}}


\maketitle

\begin{abstract}
	Satellite-terrestrial networks (STNs) have emerged as a promising architecture for providing seamless wireless coverage and connectivity for multiple users. However, potential malicious eavesdroppers pose a serious threat to the private information via STNs due to their non-cooperative behavior and ability to launch intelligent attacks. To address this challenge, we propose a cognitive secure communication framework driven by multiple agents that coordinates spectrum scheduling and protection through real-time sensing, thereby disrupting the judgment of eavesdroppers while preserving reliable data transmission. On this basis, we formulate an optimization problem to maximize the secrecy probability of legitimate users, subject to a reliable transmission probability threshold. To tackle this problem, we propose a two-layer coordinated defense system. First, we develop a foundation layer based on multi-agent coordination schedule to determine the satellite operation matrix and the frequency slot occupation matrices, aiming to mitigate spectrum congestion and enhance transmission reliability. Then, we exploit generative adversarial networks to produce adversarial matrices, and employ learning-aided power control to set real and adversarial signal powers for protection layer, which actively degrades the inference capability of eavesdroppers. Simulation results demonstrate that the proposed method outperforms benchmark methods in terms of enhancing security performance and reducing power overhead for STNs in the cognitive secure communication scenario.
\end{abstract}


\section{Introduction}
\IEEEPARstart{N}{ext}-generation wireless networks spanning 5G and 6G will be required to support a wide range of applications and to accommodate sustained and often exponential growth in mobile data usage. In this context, providing differentiated connectivity alongside stable and high-quality communication becomes a key challenge\cite{b1,b2}. Meanwhile, remote and rural areas continue to be underserved due to the high cost and difficulty of deploying conventional terrestrial infrastructure. To address these issues, satellite-terrestrial networks (STNs) have emerged as a prominent architectural solution for next-generation wireless networks\cite{b3,b4}. STNs combine satellite systems with terrestrial cellular networks to provide global coverage in near real time, on-demand access, and high throughput. Moreover, their multi-layer, heterogeneous design introduces path diversity and complementary links, which can enhance reliability and service availability\cite{b5,b6}.

Although STNs can deliver high-rate and ubiquitous connectivity, there are still several challenges to overcome. Firstly, data traffic spanning the satellite and terrestrial tiers relies on the shared spectrum and power budgets, leading to co-channel interference within and between tiers and degrading the quality of service\cite{b7}. Secondly, the broadcast nature of radio signals and the openness of satellite and terrestrial channels make secure communication a critical issue\cite{b8}. With rapid advances in attack and counter-attack techniques, modern intelligent jamming and eavesdropping devices can precisely probe and suppress 1-6 GHz frequency-hopping and mixed narrowband and wideband transmissions, thereby increasing the difficulty of defense\cite{b9}. In addition, public incident reports also indicate that billions of privacy records have been exposed worldwide in recent years\cite{b10}. Therefore, STNs require innovative methods to achieve efficient data delivery and secure, cognitive communication in heterogeneous multi-tier environments.

Dynamic spectrum control (DSC) is widely regarded as an effective method of enabling secure communications in large-scale, high-load settings, such as STNs\cite{b11}. The key concept involves managing scarce spectrum and power resources through precise coordination and proactive planning, so that scheduling decisions can meet service objectives while mitigating interference and malicious jamming\cite{b12}. Moreover, by dynamically reshaping the available time-frequency resources, legitimate users transmit over time-varying slot combinations, improving robustness against passive eavesdropping. Classical DSC relies on a control matrix, which is typically distributed to nodes and their associated users prior to each transmission window and remains fixed for a given period. Without cryptographic protection, once such a matrix is inferred or leaked, an adversary can track and replay the information at low cost by monitoring the occupied slots aggressively. In view of the rapid progress in artificial intelligence (AI), deep reinforcement learning (DRL) can learn state-dependent optimal policies and make verifiable decisions under prescribed service targets, generative adversarial networks (GANs) can synthesize structure-consistent samples that match the training set, and these have proved effective in various communication and sensing tasks\cite{b13,b14}. By embedding DRL in the DSC framework replaces a fixed control matrix with a policy-driven, time-varying slot occupation, we can take scheduling actions step by step from real-time observations. Meanwhile, GANs can be leveraged to generate adversarial patterns that carry no useful payload. These patterns can then be processed in parallel via orthogonal resources to prevent mutual interference. Consequently, even a learning-enabled eavesdropper will experience perturbed decision boundaries, resulting in detection and decoding errors.

Motivated by the above analysis, we propose a cognitive secure communication method via multi-agent DRL (MADRL) and GANs, which achieves orderly link scheduling and secure transmission against eavesdroppers with learning capabilities and precise attack capabilities. Based on the constructed network models, we formalize an optimization problem designed to enhance network security while ensuring transmission reliability. To solve this problem, we develop a multi-agent decision-making framework to create a matrix that controls the operating frequency band of all satellites and a series of matrices that control the frequency slot occupation of users served by all nodes. We then adopt GANs to generate adversarial matrices similar to the real scheduling matrices. In addition, we further introduce multi-agent learning modules for the optimal decisions of power allocation, which are trained by a group of deep double Q-networks (DDQN) to improve security performance and reduce additional power overhead. The main contributions are summarized as follows:
\vspace{-1pt}
\begin{itemize}
	\item We propose a method of multi-agent coordinated link scheduling and spectrum control for cognitive secure transmission in STNs. Taking real-world transmission scenarios into account, we derive the closed-form expressions of secrecy probability (SP) and reliable transmission probability (RTP). Based on these expressions, we formulate an optimization problem that aims to maximize SP while maintaining a high threshold of RTP. Subsequently, we develop a two-layer coordinated framework to jointly optimize spectrum occupation and adversarial defense.
	
	\item We employ multi-agent coordination schedule to determine the matrices for satellite operations and frequency slot occupation, guiding the data of legitimate users from different node to occupy different frequency slots for non-interfering transmission. Real-time spectrum sensing can continuously update the state of the network environment, and the reward function encompasses both security, reliability, and interference avoidance. We adopt the DDQN to train these agents, which provides a foundation layer of defense for the proposed transmission method.
	
	\item We leverage the GANs to generate adversarial matrices that are similar to the real scheduling matrices. Moreover, we incorporate additional generator loss functions to align the structures and statistics of the adversarial and legitimate transmission patterns more closely. We further allocate transmit power for the real and adversarial signals based on learning-aided cooperative power control, which aims to maximize confusion for eavesdroppers under minimal overhead without affecting legitimate transmissions, thereby forming an effective protection layer.
	
	\item We present the results of extensive simulations to verify the effectiveness of our method for STNs, and demonstrate the superiority of the proposed transmission method in terms of enhancing network security and reducing additional power overhead compared with other existing methods in the cognitive secure communication scenario.
\end{itemize}

The rest of this paper is arranged as follows. Section \uppercase\expandafter{\romannumeral2} reviews the related works. The system model is presented in Section \uppercase\expandafter{\romannumeral3}. Section \uppercase\expandafter{\romannumeral4} formulates the optimization problem. The proposed multi-agent-driven method is described in Section \uppercase\expandafter{\romannumeral5}. Section \uppercase\expandafter{\romannumeral6} shows the simulation results. Finally, the conclusions and future directions are drawn in Section \uppercase\expandafter{\romannumeral7}.

\section{Related Work}
In this section, we provide a brief review of related work on secure transmission methods and AI-based secure communication solutions to highlight the motivation of our own work.

The security of STNs has garnered increasing attention, and one effective approach is to involve reconfigurable intelligent surfaces (RISs)\cite{b15,b16}. Niu \textit{et al.} proposed an active RIS-assisted method to enhance security and minimize interference\cite{b15}. Building on this idea, Ge \textit{et al.} further improved security by employing cooperative beamforming and addressing imperfect channel state information (CSI)\cite{b16}. Another promising direction is unmanned aerial vehicle (UAV)-assisted secure relaying\cite{b17,b18,b19}. Sharma \textit{et al.} introduced a 3D mobile UAV relaying method based on decode-and-forward principles and analyzed the secrecy performance under different eavesdropper locations\cite{b17}. Subsequently, an aerial bridge scheme proposed by Wang \textit{et al.} established secure tunnels via UAVs, effectively preventing data leakage without disrupting legitimate traffic\cite{b18}. Later, He \textit{et al.} extended these efforts by integrating UAVs with low-resolution ADC/DAC architectures\cite{b19}. Apart from RIS and UAV techniques, beamforming and efficiency optimization methods have also been widely studied\cite{b20,b21,b22}. In 2021, a robust beamforming method for STNs successfully enhanced network rates while meeting secrecy constraints\cite{b20}. In parallel, Lin \textit{et al.} developed an efficient hybrid beamforming method, optimizing beamformers to improve secrecy-energy efficiency\cite{b21}. Furthermore, Cao \textit{et al.} explored relay-user pairing methods and designed optimal allocation based on different CSI levels to enhance physical-layer security\cite{b22}. Although these studies offer diverse and complementary approaches, they typically assume static adversaries.

Recent advances in AI-driven methods have substantially elevated the protection of wireless networks\cite{b23,b24,b25,b26,b27,b28,b29,b30}. Early explorations by Miller \textit{et al.} surveyed the robustness of neural networks against adversarial attacks and compared different defense techniques\cite{b23}. In response to growing privacy concerns, Chen \textit{et al.} formulated secure deep learning (DL) protocols for the Internet of Things (IoT), mitigating attacks based on GANs and reducing data leakage\cite{b24}. Complementing these efforts, Wang \textit{et al.} shifted the research focus towards industrial IoT, where they pioneered DRL-based attack tolerance approaches, further augmented by GANs-generated traffic for defensive resilience\cite{b25}. Lin \textit{et al.} advanced this field by employing a joint DRL and GANs framework, which aims to optimize resource management in cognitive IoT systems\cite{b26}. Expanding the scope of applications, Yao \textit{et al.} contributed collaborative GANs-enhanced solutions for privacy-preserving intrusion detection\cite{b27}. Meanwhile, Wen \textit{et al.} addressed the dual challenges of intelligent jamming and covert communication by leveraging GANs-optimized beamforming and power allocation to improve physical layer security\cite{b28}. As generative AI models continue to proliferate, Zhao \textit{et al.} provided a comprehensive perspective on the effect of GANs in facilitating security and reliability in complex environments\cite{b29}. In the sphere of decentralized federated learning (DFL), Xu \textit{et al.} designed difference-weight transmission mechanisms to combat eavesdropping\cite{b30}. However, they are not adapted to the unique features and dynamic conditions of STNs.

Different from the existing studies, we aim to develop a multi-agent-driven solution for cognitive secure in STNs. With the emergence of advanced jamming and attack technologies, adversaries may be able to infer transmission patterns and launch targeted attacks. This presents a significant challenge, as it requires the implementation of an active cognitive defense method. To the best of our knowledge, no existing work has considered network security and information transmission in STNs under the threat from eavesdroppers with intelligent attack capabilities. However, only by considering these factors holistically can STNs meet the security and reliability demands of real-world applications. This gap motivates our work. In this study, we optimize link scheduling and spectrum control jointly to proactively counter eavesdropping and achieve cognitive communication in STNs. In the following, we present the detailed design and analysis of our proposed communication method, which aims to handle learning-enabled eavesdropping and maintain optimal network performance in STNs.

\section{System Model}
\subsection{{Network Description}} 
\begin{figure}[t!]
	\centerline{\includegraphics[width=1\linewidth]{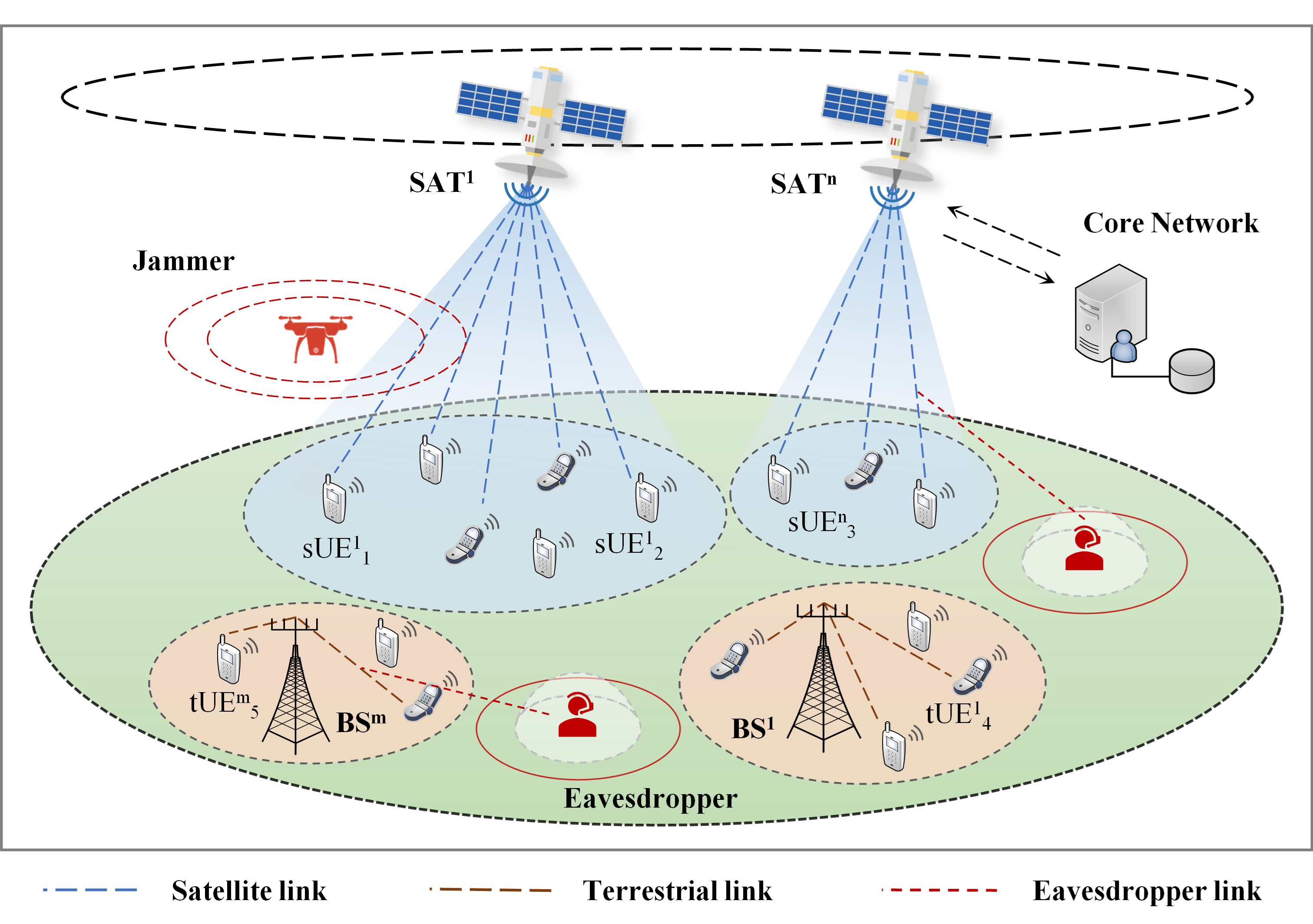}}
	\setlength{\belowcaptionskip}{-0.5cm}   
	\caption{Illustration of the STNs. SAT$^1$: LEO satellite $1$; SAT$^n$: LEO satellite $n$; BS$^1$: base station $1$; BS$^m$: base station $m$; sUE$^1_1$: the legitimate UE $1$ associated with the SAT$^1$; sUE$^1_2$: the legitimate UE $2$ associated with the SAT$^1$; sUE$^n_3$: the legitimate UE $3$ associated with the SAT$^n$; tUE$^1_4$: the legitimate UE $4$ associated with the BS$^1$; tUE$^m_5$: the legitimate UE $5$ associated with the BS$^m$.}
	\label{f1}
\end{figure}

In line with the practical deployments and common STNs models\cite{b19,b20,b21,b22}, a typical STNs scenario is considered in this work as shown in Fig.~\ref{f1}. It consists of low Earth orbit (LEO) satellites (SATs), ground base stations (BSs), legitimate user equipment (UEs), and adversaries in the form of malicious jammers and illegal eavesdroppers. Integration at the system level between the satellite and terrestrial layers enables seamless and ubiquitous connectivity over wide areas, ensuring continuous and high-quality service for highly mobile UEs. In the considered system, traffic originates from internet content or service providers and will be received by all legitimate UEs. The core network (CN) coordinates packet scheduling and forwarding, which also maintains attachment state and reachability of each UE and selects the best forwarding path.

Motivated by its nominal capacity of around 340 million tasks per day, the Geely constellation, a representative LEO satellite system, is considered in this work. Its first deployment phase has been completed to date, with 64 satellites in orbit providing near-global coverage, excluding the polar regions\cite{b31}. Due to the regular and predictable motion of these satellites, a discrete-time model is adopted to capture link dynamics. Thus, the communication timeline is divided into short slots and the positions of the satellites are treated as fixed within each slot. Let the set of LEO satellites at a fixed altitude be $\mathcal{N_S}=\left\{\rm{SAT}^1,\rm{SAT}^2,\ldots,\rm{SAT}^N\right\}$, where $N$ is the total number of LEO satellites. These form the macro layer of the STNs and provide reliable connectivity to legitimate UEs with a higher transmit power of $p_s$ (from 50 to 200 W). In contrast, the terrestrial segment comprises a set of M BSs $\mathcal{N_B}=\left\{\rm{BS}^1,\rm{BS}^2,\ldots,\rm{BS}^M\right\}$, all of which are deployed as macro-stations within the STNs. These stations have a small coverage radius and form multiple non-overlapping macrocells within the beam footprints of the LEO satellite constellation. As supplementary deployment facilities handling ever-growing traffic demands and achieving seamless coverage, they only serve the legitimate UEs within their communication macro-cells at a low transmit power level $p_b$ (from 5 to 40 W). Both the SATs and BSs are equipped with uniform linear arrays (ULAs) to support secure and efficient multi-user access. 

In this work, cognitive communication refers to a system-level modelling assumption that is driven by real-time sensing. At each time slot, SATs and BSs perform wideband spectrum sensing and link probing over their footprints to obtain the environment state, including slot occupation, interference distribution, and basic link-quality indicators. Subsequently, the CN combines these multi-source observations to construct the network-wide time-varying spectrum maps and interference heatmaps, thereby yielding a global view for decision-making.

To elaborate, we focus on the downlink procedures in STNs. For satellite-side UEs, the SAT operates in transparent relay mode. It receives the service signal and forwards it to the coverage area, where the satellite terminal receives and decodes the data. For terrestrial-side UEs, the BS transmits on cellular bands and UEs perform reception and decoding. Within each time slot, the number of UEs and their associations are fixed. As time passes and locations change, UEs requiring ubiquitous coverage tend to prefer satellite access, whereas those seeking higher capacity are more likely to connect to BSs. Hence, in each time slot, a terminal selects either a SAT or a BS according to its position and channel state, with the aim of achieving continuous and stable service without needing to scheduling details. It should be noted that the satellite and terrestrial downlinks operate on their independent transmission timing domains. Let $\mathcal{K}=\left\{{s{\rm{U}}{{\rm{E}}_1},\ldots,s{\rm{U}}{{\rm{E}}_A},t{\rm{U}}{{\rm{E}}_1},\ldots,t{\rm{U}}{{\rm{E}}_B}} \right\}$ denote the set of legitimate UEs in STNs, where $A$ and $B$ are positive integers representing the number of UEs attached to the satellite and terrestrial links, respectively. The total number of UEs denotes $K=A+B$ and is treated as constant.

As shown in Fig.~\ref{f1}, the network is exposed to two main types of adversary: jammers and eavesdroppers. As the radio spectrum is limited and scarce, malicious jammers may transmit either intermittent or continuous interference over shared bands, which degrades the quality of service and can even cause increased error rates or service interruption in severe cases. Eavesdroppers, the main adversaries threatening network security, can be randomly distributed in STNs and desire to intercept privacy traffic transmitted by legitimate UEs with the assistance of an energy detector. In this work, illegal eavesdroppers are modeled as non-cooperative, learning-enabled agents who can observe the environment to predict likely frequency slot occupation and then perform a targeted binary hypothesis test on these slots. This model more accurately reflects practical threat conditions and therefore imposes stricter security requirements on STNs. Moreover, although the integrated CN can orchestrate satellite and terrestrial links, the absence of proactive spectrum planning and fine-grained management across these two segments can lead to inter-tier interference. In addition, rising real-time traffic and concurrent networks sharing the same spectrum bands may also introduce further co-channel interference and inter-link interference.

\subsection{{Channel Model}}
In this work, we employ discrete-time modeling to quantify the time-varying characteristics of the maneuverable LEO constellation network by dividing the communication timeline into multiple short time slots. Without loss of generality, we assume that all legitimate UEs in STNs transmit their data simultaneously over a discrete-time channel comprising $p$ time slots ($\mathcal{T}=t_1, t_2, \ldots, t_p$), where $p$ is a positive integer denoting the total number of divided time slots, and $t_p$ is the $p$-th time slot. We further assume that each legitimate UE needs to occupy $L\left(L \leq p\right)$ consecutive time slots within one transmission period to deliver a data block. Here, $L$ denotes the number of consecutive time slots required to meet the specified throughput and delay targets during this period. Relying on their respective service quality requirements and business scale targets, legitimate UEs may repeat the above communication procedure over multiple periods for their purposes.

\begin{figure}[t!]
	\centerline{\includegraphics[width=1\linewidth]{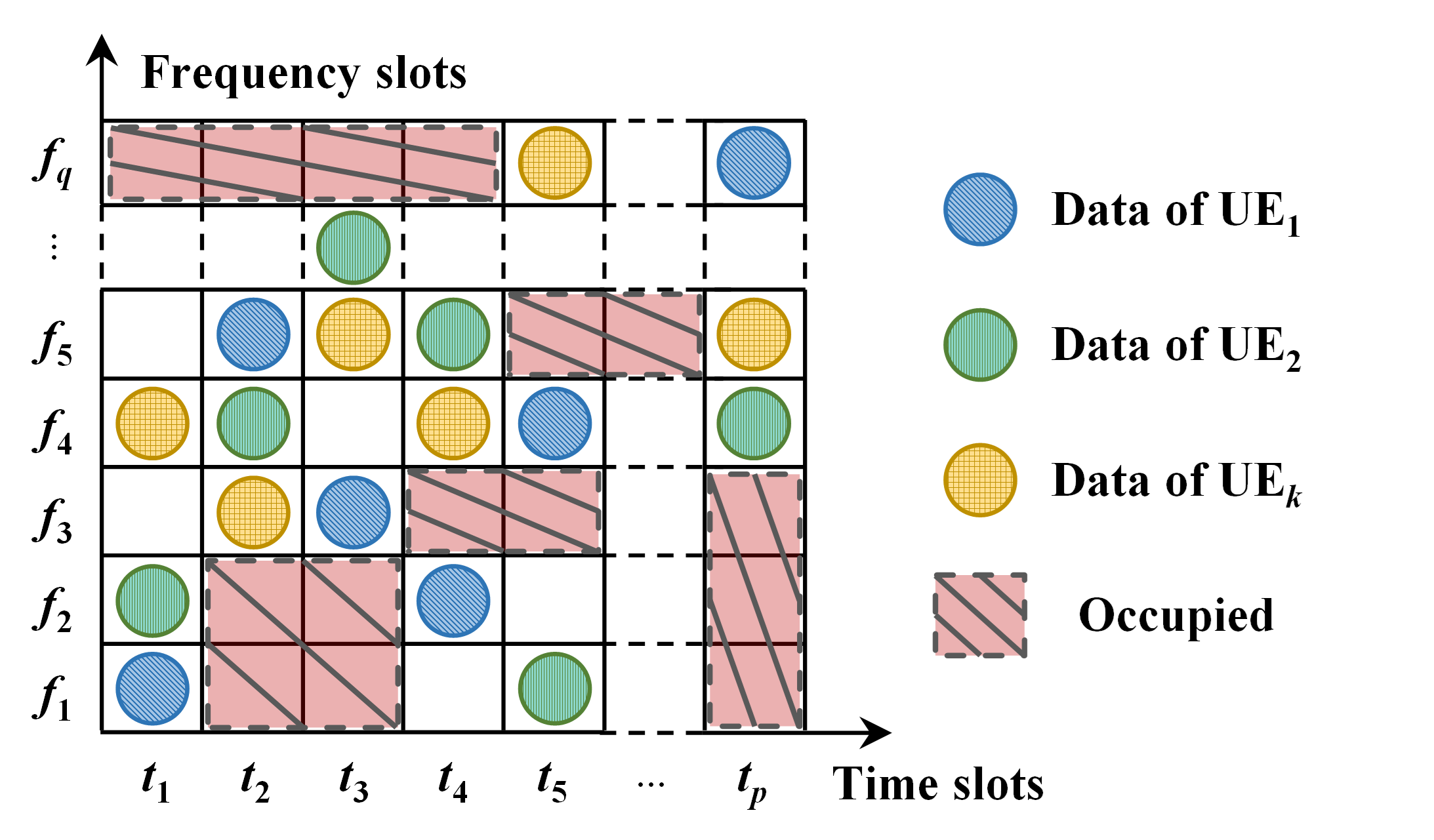}}
	\setlength{\belowcaptionskip}{-0.5cm}   
	\caption{Illustration of the communication channel. UE$_1$: the legitimate UE $1$; UE$_2$: the legitimate UE $2$; UE$_k$: the legitimate UE $k$; and they are associated with the same node.}
	\label{f2}
\end{figure}

To accurately capture the communication process in STNs, the frequency dimension must be introduced alongside time to represent the available spectrum resources. For efficient and coordinated utilization of scarce resource while accommodating the growing high-concurrency and multi-user demand, the shared transmission bandwidth is divided into $q\left(q  >  K\right)$ mutually non-overlapping frequency slots in advance, which form the set of frequency slots $\mathcal{F} = \left\{{ f_1, f_2,\ldots, f_q }\right\}$, where $q$ is a positive integer denoting the total number of the divided frequency slots, and $f_q$ denotes the $q$-th frequency slot. Combining the discretization of the time and frequency dimensions yields the time-frequency domain channel model shown in Fig.~\ref{f2}. Legitimate UEs can dynamically select different frequency slots for the entire period based on transmission schemes agreed upon with the network nodes to accomplish information communication. Since these frequency slots may also be targeted by eavesdroppers for monitoring purposes, the aforementioned design reduces the probability of full message recovery. Even if eavesdroppers obtain fragmented data by observing partial frequency slots, they would still struggle to reconstruct the complete transmission payload. For illustration, UE$_1$ in Fig.~\ref{f2} occupies slot $f_1$ at time $t_1$ and switches to slot $f_5$ at time $t_2$ for continued transmission, and so on. The resource allocation and access for all UEs in Fig.~\ref{f2} are centrally coordinated by a single control node (a SAT or a BS), so that stable and reliable delivery over this time-frequency grid can be maintained. Note that $\mathcal{F}$ can also be shared by all nodes in STNs. SATs conduct operations simultaneously in the downlink frequency band, and BSs also operate in the same spectrum band. Clearly, the interference between the satellite and terrestrial segments can be mitigated by core-level coordination, whereas inter-cell coupling is more difficult to avoid. Furthermore, a malicious jammer may inject energy into these frequency slots and further degrade the link environment.

In our work, we assume that the SAT remains stable in its work orbit for a short period and can cover all UEs. This scenario can be considered a short-term fading environment for a satellite-to-ground downlink under clear line-of-sight (LoS) conditions. Based on the slot-level modeling, we ignore the impact of large-scale shadow fading on the link. Comparably, small-scale fading, path loss, and Doppler shift are explicitly accounted for\cite{b32}. The small-scale effect is captured by a Rician model. Path loss is incorporated separately via the Friis transmission relation. Additionally, the Doppler term induces phase rotation and frequency offset in the LoS component. As a result, the representation of the satellite channel is as follows
\begin{equation}
{h_{\rm{s}}}\left( t \right) = \sqrt {\frac{\Omega}{{\Omega + 1}}} {h_{{\rm{LoS}}}}\left( t \right) + \sqrt {\frac{1}{{\Omega + 1}}} {h_{{\rm{scatter}}}}\left( t \right),
	\label{e1}
\end{equation}
where $\Omega$ is Rician factor denoting power ratio of LoS to scatter, ${h_{{\rm{LoS}}}}\left( t \right) = {e^{j\left( {2\pi {f_D}t + {\phi _0}} \right)}}$ is the deterministic unit-amplitude component with Doppler $f_D$, and ${h_{{\rm{scatter}}}}\left( t \right) \sim \mathcal{CN}\left( {0,1} \right)$ denotes the zero-mean Gaussian scattering component. Writing ${h_{{\rm{scatter}}}}\left( t \right)=x + jy$ with $x \sim \mathcal{N}\left( {0,\frac{1}{2}} \right)$, $y \sim \mathcal{N}\left( {0,\frac{1}{2}} \right)$, and $x \bot y$. Therefore, ${h_{\rm{s}}}\left( t \right)$ is a complex Gaussian with non-zero mean and the instantaneous power gain ${\left| {{h_{\rm{s}}}\left( t \right)} \right|^2}$ follows a non-central chi-squared distribution with two degrees of freedom and non-centrality parameter $\lambda  = 2\Omega$. The probability density function (PDF) of the power gain ${\left| {{h_{\rm{s}}}\left( t \right)} \right|^2}$ is given by
\begin{equation}
{f_{{G_{\rm{s}}}}}\!\left( g \right) \!=\! \left( {\Omega \!+\! 1} \right){e^{ \!-\! \Omega \!-\! \left( {\Omega \!+\! 1} \right)g}}{I_0}\left( {\!2\sqrt {\Omega\left( {\Omega \!+\! 1} \right)g} } \right)\!,{\rm{ }}g \!\ge\! 0,
		\label{e2}
\end{equation}
where ${I_0}\left(  \cdot  \right)$ is the zero-order modified Bessel function.

The wireless channel from a terrestrial BS to a UE is modeled under non-line-of-sight (NLoS) conditions\cite{b33}. Under rich scattering and a narrowband assumption, the complex envelope is a zero-mean circularly symmetric, complex Gaussian process corresponding to Rayleigh flat fading. Accordingly, we assume that each terrestrial link in STNs is statistically independent, experiences Rayleigh flat fading and has an exponential distribution for its instantaneous power gain. Specifically, the value of the channel power gain is ${\left| {{h_{\rm{su}}}} \right|^2}$ between the SAT and legitimate UEs, and ${\left| {{h_{\rm{se}}}} \right|^2}$ between the SAT and eavesdroppers, and ${\left| {{h_{\rm{tu}}}} \right|^2}$ between the BS and legitimate UEs, and ${\left| {{h_{\rm{te}}}} \right|^2}$ between the BS and eavesdroppers.

\section{Problem Description}
In this section, we formalize the problem of secure and efficient STNs. First, we outline the main idea of the optimization problem. Then, we define the decision variables and the optimization objectives. Finally, we present the formal problem statement and prove that it is an NP-hard problem.

\vspace{-0.2cm}
\subsection{Problem Statement}
While the STNs prioritize secure transmission, this must not result in a significant loss of communication performance. Consequently, our objective is to enhance the achievable security performance while maintaining high reliability. We assume that the eavesdropper can monitor multiple frequency slots over a period of time, who is equipped with a learning-enabled device to observe and predict frequency slot occupation for launching targeted attacks. To prevent this threat, our overarching objective is to schedule transmission links for all network nodes in both the time and frequency domains, while performing proactive frequency slot allocation for each UE, thereby optimizing system-level security performance. Firstly, the integrated CN separates and orchestrates the operating bands of multiple satellites, which can suppress inter-satellite interference among legitimate terminals served by different satellites. Secondly, on the constructed time-frequency grid channel, the SATs and BSs assign mutually distinct transmission frequency slots to the terminals they serve, which helps to evade eavesdropping detection and reduces the impact of malicious jamming. These two measures act in concert to enable conflict-free scheduling while meeting security constraints and preserving the stability of data delivery. It is important to highlight that, although the ideal goal is to improve security performance and transmission quality simultaneously, in practice, one can first guarantee the baseline reliability and stability of transmission, and then maximize the security performance on this basis. This can equally achieve the desired equilibrium between the network security and performance of STNs.

\subsection{Decision Variables}
Let one transmission period in the network consist of $L$ time slots ($\mathcal{L}=t_1, t_2, \ldots, t_L$), and let the global set of spectrum partitions be denoted by $\mathcal{F} = \left\{{ f_1, f_2,\ldots, f_q }\right\}$, and the index be denoted by $\mathcal{Q} = \left\{{ 1, 2,\ldots, q }\right\}$. The set of all network nodes is $\mathcal{Z}= \mathcal{N_S} \cup \mathcal{N_B}$, where $\mathcal{N_S}$ is the SAT set and $\mathcal{N_B}$ is the BS set. The set of legitimate UE served by $z \in \mathcal{Z}$ is denoted by $\mathcal{K_Z} = \left\{{ 1, 2,\ldots, K_z }\right\}$. Based on the above analysis, the following decision variables need to be determined jointly: \textit{i)} a satellite band operation matrix $\mathbf{S}= \left\{ {{S_{n,j}}} \right\} \in \left\{ {0,1} \right\},\forall n \in \mathcal{N_S},j \in \mathcal{Q}$, where $\left\{ {{S_{n,j}}} \right\} = 1$ means that SAT$^n$ operates on frequency slot $f_j$; \textit{ii)} frequency slot occupation matrices of each node $\mathbf{X}=\left\{ {{X_{z,l,k,f}}} \right\} \in \left\{ {0,1} \right\}$ satisfying $\sum\nolimits_{f \in \mathcal{F}} {{X_{z,l,k,f}} = 1,} \forall z \in \mathcal{Z},l \in \mathcal{L},k \in \mathcal{K_Z}$, which specifies the unique frequency slot chosen by node $z$ for UE $k$ in time slot $t_l$; \textit{iii)} adversarial spectrum matrices of each node $\mathbf{A}=\left\{ {{A_{z,l,k,f}}} \right\} \in \left\{ {0,1} \right\}$, where $\sum\nolimits_{f \in \mathcal{F}} {{A_{z,l,k,f}} = 1,} \forall z,l,k$, and there is no overlap between $\mathbf{A}$ and $\mathbf{X}$ at the same $(z,l,k,f)$; and \textit{iv)} transmit power control matrices $\mathbf{p}=\left\{ {{p_{z,l,k,f}}} \right\}$ of all network node for legitimate data and adversarial signals at each time-frequency position. In this matrix, ${p_{z,l,k,f}^d}$ denotes the power for legitimate data, and ${p_{z,l,k,f}^a}$ denotes the power for adversarial signaling, and they satisfy $\sum\nolimits_k {\big( {p_{z,l,k,f}^d + p_{z,l,k,f}^a} \big)}  \le {p_z}, \forall z \in \mathcal{Z},l \in \mathcal{L},k \in \mathcal{K_Z}$, where $p_z$ is the maximum transmit power of node $z$. These variables describe fine-grained occupation and power decisions on the time-frequency grid and are suitable for joint optimization with network security and performance objectives.

\subsection{Optimization Objectives} 
Our primary goal is to minimize the probability of the eavesdropper successfully identifying and decoding legitimate data in STNs, \textit{i.e.}, to maximize the probability that the eavesdropper will fail to obtain useful information. Accordingly, we proceed with the following analysis.

\subsubsection{Satellite links}
We assume that the eavesdropper has learning capabilities and can predict which frequency slots may carry legitimate traffic. However, it still relies on energy detection to determine whether data is actually being transmitted. Let the event of hitting the target frequency slot be denoted as $H_1$, which is uniquely determined by the decision variable $\mathbf{X}$. Then the signal-to-interference-plus-noise ratio (SINR) for the eavesdropper in this frequency slot is
\begin{equation}
	\gamma _{n,l,k}^{\left( e \right)} = \frac{{p_{n,l,k}^dg_{n,l,k}^{\left( e \right)}}}{{I_{n,l}^{{\rm{inter,}}\left( e \right)}\left( {{\bf{S}},{\bf{X}},{\bf{A}}} \right) + I_{n,l}^{{\rm{intra,}}\left( e \right)}\left( {{\bf{X}},{\bf{A}}} \right) + {N_0}}},
	\label{e3}
\end{equation}
where $p_{n,l,k}^d$ is the transmit power for legitimate data on the frequency slot, and $g_{n,l,k}^{\left( e \right)} = {\left| {{h_{\rm{se}}}\left( {{f_j}} \right)} \right|^2}$ is the channel gain between the SAT$^n$ and the eavesdropper at time $t_l$. According to the channel modeling, $g_{n,l,k}^{\left( e \right)}$ follows a non-central chi-squared distribution with two degrees of freedom and non-centrality parameter ${\lambda _e}\left(\lambda _e=2\Omega_e\right)$. Moreover, ${I_{n,l}^{{\rm{inter,}}\left( e \right)}}$ denotes inter-satellite interference, and ${I_{n,l}^{{\rm{intra,}}\left( e \right)}}$ denotes intra-satellite co-channel interference. Due to spectrum band orchestration by the CN and orthogonal scheduling of the frequency slot occupation, both terms can be eliminated by design. If the target frequency slot is missed (event $H_0$), \textit{i.e.}, the monitored frequency slot carries no data, the observed SINR is given by
\begin{equation}
	\gamma _{n,l,k}^{\left( e \right)} = \frac{{p_{n,l,k}^ag_{n,l,k}^{\left( e \right)}}}{{ {N_0}}},
	\label{e4}
\end{equation}
where $p_{n,l,k}^a$ is the adversarial power. To induce detection errors, the detection threshold ${\tau _e}$ is chosen such that $\gamma _{n,l,k}^{\left( e \right)}<\tau _e$ under $H_1$ while $\gamma _{n,l,k}^{\left( e \right)}>\tau _e$ under $H_0$. If a uniform frequency slot randomization is adopted to flatten observable patterns, the selection probability of an occupied frequency slot is $\Pr \left( {{f_j}} \right) = {1 \mathord{\left/{\vphantom {1 q}} \right.\kern-\nulldelimiterspace} q}$, yielding $\Pr \left( {{H_1}} \right) = {{{1}} \mathord{\left/{\vphantom {{{1}} q}} \right.\kern-\nulldelimiterspace} q}$ and $\Pr \left( {{H_0}} \right) = 1-{{{1}} \mathord{\left/{\vphantom {{{1}} q}} \right.\kern-\nulldelimiterspace} q}$. Consequently, we define the probability that valid information cannot be obtained by the eavesdropper as the secrecy probability (SP), which can be expressed as
\begin{equation}
	\begin{aligned}
		&{P_{s,e}}\left( {{\bf{S}},{\bf{X}},{\bf{A}},{\bf{p}}} \right)\\
		&= \underbrace {\Pr \!\left( {{H_1}} \right)\!\Pr \left( {\gamma _{n,l,k}^{\left( e \right)} \!<\! {\tau _e}} \right)}_{{I_a}} + \underbrace {\Pr\! \left( {{H_0}} \right)\!\Pr \left( {\gamma _{n,l,k}^{\left( e \right)} \!>\! {\tau _e}} \right)}_{{I_b}}.
	\end{aligned}
	\label{e5}
\end{equation}
Based on the PDF of $g_{n,l,k}^{\left( e \right)}$ and the change of variable $v\! =\! \sqrt {2\left( {{\Omega_e} \!+\! 1} \right)g} $, the first term $I_a$ of Eq.~(\ref{e5}) can be written as
\begin{equation}
	\begin{aligned}
		{I_a}&={\frac{{{1}}}{q}\!\!\int_0^{\sqrt {\frac{{{2\left( {{\Omega_e} \!+\! 1} \right)}\!{N_0}\!{\tau _e}}}{{p_{n,l,k}^d}}} } \!\!{\exp\! \left(\! { - \frac{{{v^2}\! +\! 2{\Omega_e}}}{2}} \!\right)\!{I_0}\!\left( \!{\sqrt {2{\Omega_e}} v} \!\right)} dv}\\
		&= \frac{{{1}}}{q}\left( {1 - {Q_1}\left( {\sqrt {2{\Omega_e}} ,\sqrt {2\left( {{\Omega_e} + 1} \right)\frac{{{N_0}}}{{p_{n,l,k}^d}}{\tau _e}} } \right)} \right),
	\end{aligned}
	\label{e7}
\end{equation}
where ${Q_1}\left( { \cdot , \cdot } \right)$ is the first-order Marcum-$Q$ function. By the same argument, the second term $I_b$ can be presented by
\begin{equation}
	{I_b} = \left( {1 - \frac{{{1}}}{q}} \right){Q_1}\left( {\sqrt {2{\Omega_e}} ,\sqrt {2\left( {{\Omega_e} + 1} \right)\frac{{{N_0}{\tau _e}}}{{p_{n,l,k}^a}}} } \right).
	\label{e8}
\end{equation}

Combining Eq.~(\ref{e7}) and (\ref{e8}) gives the SP for UEs associated with SAT$^n$. Note that ${1 \mathord{\left/{\vphantom {{1} q}} \right.\kern-\nulldelimiterspace} q}$ clearly models the parallel monitoring capability of eavesdropper. If the listening set or frequency slot occupation is non-uniform, this weight must be adjusted. A larger SP means a lower chance of successful decoding by the eavesdropper, which aligns with the security objective.

\subsubsection{Terrestrial links}
When the eavesdropper monitors the terrestrial segment and attacks the frequency slot carrying legitimate data, the instantaneous SINR is given by
\begin{equation}
\gamma _{m,l,k}^{\left( e \right)} = \frac{{p_{m,l,k}^dg_{m,l,k}^{\left( e \right)}}}{{I_{m,l}^{{\rm{inter}},\left( e \right)}\left( {{\bf{X}},{\bf{A}}} \right) + I_{m,l}^{{\rm{intra}},\left( e \right)}\left( {{\bf{X}},{\bf{A}}} \right) + {N_0}}},
	\label{e9}
\end{equation}
where $p_{m,l,k}^d$ is the power for legitimate signal on this frequency slot, and $g_{m,l,k}^{\left( e \right)} = {\left| {{h_{\rm{te}}}\left( {{f_j}} \right)} \right|^2}$ denotes the channel gain between the BS$^m$ and the legitimate UE, which follows an exponential distribution with parameter ${\omega _e}$. The terms ${I_{m,l}^{{\rm{inter}},\left( e \right)}\left( {{\bf{X}},{\bf{A}}} \right)}$ and ${I_{m,l}^{{\rm{intra}},\left( e \right)}\left( {{\bf{X}},{\bf{A}}} \right)}$ represent inter-cell and intra-cell co-channel interference in the terrestrial segment, respectively, and can be set to zero under ideal orthogonal scheduling. If the eavesdropper inspects an idle frequency slot, the observed SINR is given by
\begin{equation}
	\gamma _{m,l,k}^{\left( e \right)} = \frac{{p_{m,l,k}^ag_{m,l,k}^{\left( e \right)}}}{{ {N_0}}},
	\label{e10}
\end{equation}
where $p_{m,l,k}^a$ is the adversarial power of BS$^m$ at time $t_l$. Hence, for a legitimate UE associated with the terrestrial BS, the SP can be expressed as
\begin{equation}
	\begin{aligned}
		&{P_{t,e}}\left( {{\bf{X}},{\bf{A}},{\bf{p}}} \right) \\
		&\!=\! \Pr \left( {{H_1}} \right)\!\Pr \!\left( {\gamma _{m,l,k}^{\left( e \right)} \!<\! {\tau _e}} \right) + \!\Pr \left( {{H_0}} \right)\!\Pr \left( {\gamma _{m,l,k}^{\left( e \right)} \!>\! {\tau _e}} \right)\\
		&\!=\! \frac{{{1}}}{q}\!\left(\! {1 \!-\! \exp \left( { - \frac{{{N_0}{\tau _e}}}{{{\omega _e}p_{m,l,k}^d}}} \!\right)} \!\right) \!+\! \left(\! {1 - \frac{{{1}}}{q}} \!\right)\exp\! \left( { - \frac{{{N_0}{\tau _e}}}{{{\omega _e}p_{m,l,k}^a}}} \right).
	\end{aligned}
	\label{e11}
\end{equation}

\begin{figure}[!t]
	\centering
	\vspace{-0.2cm}  
	\setlength{\abovecaptionskip}{-0.25cm}   
	\setlength{\belowcaptionskip}{-0.5cm}   
	\begin{minipage}{0.493\linewidth}
		\centering
		\includegraphics[width=\linewidth]{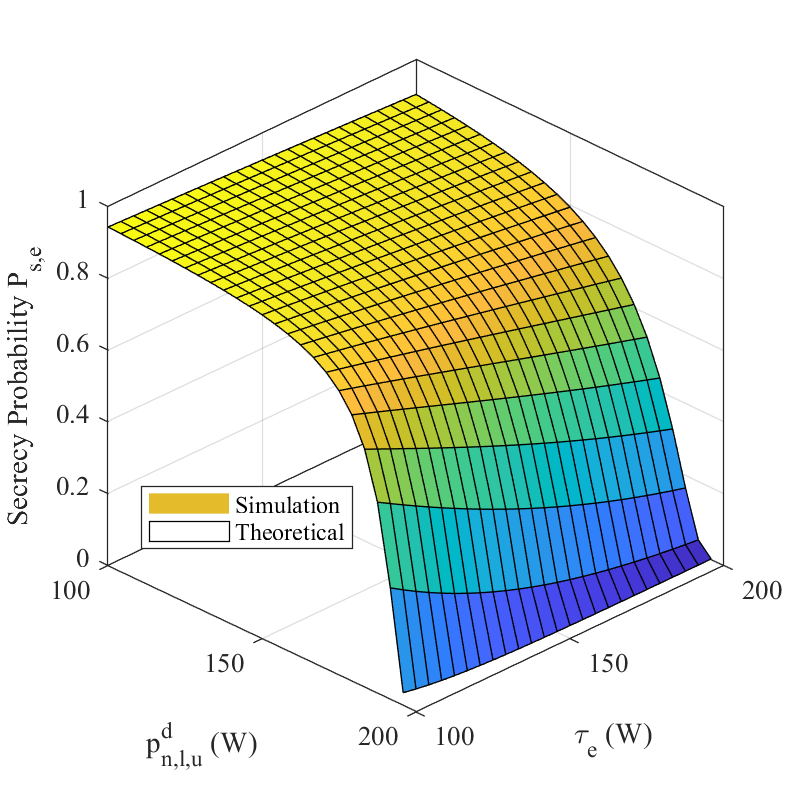}
		\parbox{\linewidth}{\centering (a)} 
		\label{f3a}
	\end{minipage}
	\begin{minipage}{0.493\linewidth}
		\centering
		\includegraphics[width=\linewidth]{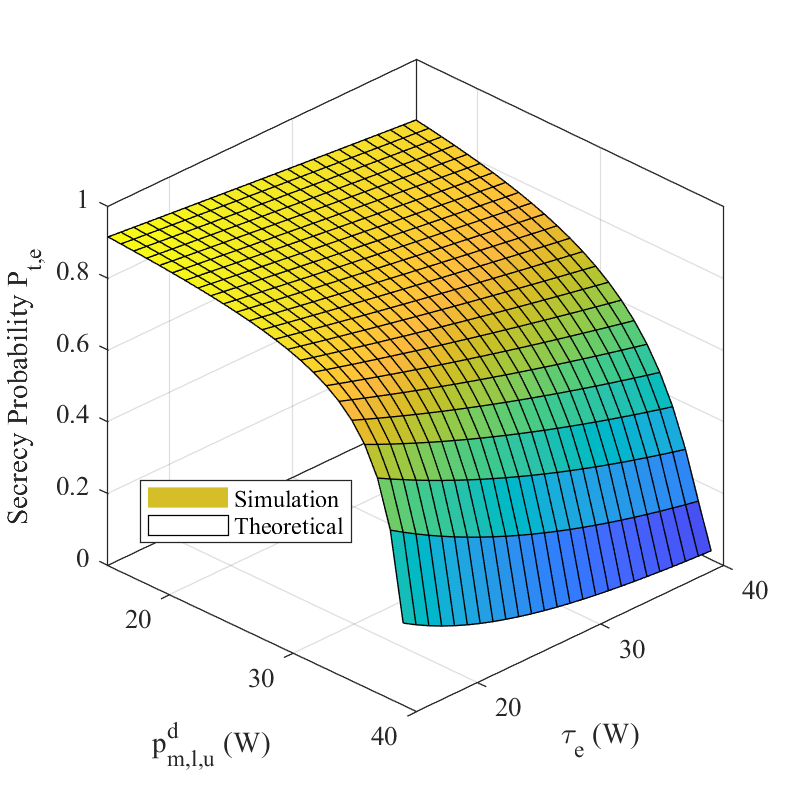}
		\parbox{\linewidth}{\centering (b)} 
		\label{f3b}
	\end{minipage}
	\caption{Comparison between the simulation and theoretical results on (a) satellite links (b) terrestrial links.}
	\label{f3}
\end{figure}

From the Fig.~\ref{f3}, we can find that the derived expressions match well with the simulation results, which confirms the correctness of our theoretical analysis. On this basis, in the subsequent framework, we adopt these theoretical results as the primary targets to formulate the overall problem.

\subsection{Main Constraints} 
When outlining the overall problem, we emphasize that, unlike multi-objective optimization, we prefer a single-objective formulation with explicit constraints. This choice is consistent with standard modeling practices and better aligns with our objective. Based on this framework, we then provide analytical derivations of the reliable transmission probability (RTP) for UEs to support the subsequent joint optimization.

\subsubsection{Satellite links}
Under the orchestration of CN and different frequency slot occupation, a satellite UE can identify the occupied slot for legitimate data in each time slot and hence perform correct decoding. For a UE served by SAT$^n$, the received SINR in time $t_l$ and frequency $f_j$ is given by
\begin{equation}
	\gamma _{n,l,k}^{\left( u \right)} = \frac{{p_{n,l,k}^dg_{n,l,k}^{\left( u \right)}}}{{I_{n,l}^{{\rm{inter}},\left( u \right)}\left( {{\bf{S}},{\bf{X}}} \right) + I_{n,l}^{{\rm{intra}},\left( u \right)}\left( {\bf{X}} \right) + {N_0}}},
	\label{e12}
\end{equation}
where $p_{n,l,k}^d$ is the transmit power for the UE, and the channel gain $g_{n,l,k}^{\left( u \right)} = {\left| {{h_{{\rm{su}}}}\left( {{f_j}} \right)} \right|^2}$ between SAT$^n$ and UE follows a non-central chi-squared law with parameter $\Omega_u$. The inter- and intra- interference, ${I_{n,l}^{{\rm{inter}},\left( u \right)}\left( {{\bf{S}},{\bf{X}}} \right)}$ and ${I_{n,l}^{{\rm{intra}},\left( u \right)}\left( {\bf{X}} \right)}$ are designed to be zero by band planning and orthogonal scheduling. Let ${{\tau _u}}$ be the decoding threshold of the UE. Then the RTP for a legitimate UE receiving from SAT$^n$ can be expressed as
\begin{equation}
	\begin{aligned}
		{P_{s,u}}\!\left( {{\bf{S}},{\bf{X}},{\bf{p}}} \right) \!&= \Pr \left( {\gamma _{n,l,k}^{\left( u \right)} > {\tau _u}} \right) \\
		&= {Q_1}\left( {\sqrt {2{\Omega _u}} ,\sqrt {2\left( {{\Omega _u} + 1} \right)\frac{{{N_0}{\tau _u}}}{{p_{n,l,k}^d}}} } \right).
	\end{aligned}
	\label{e13}
\end{equation}
where ${Q_1}\left( { \cdot , \cdot } \right)$ is the first-order Marcum-$Q$ function.

\subsubsection{Terrestrial links}
Similarly, when a legitimate UE receives data from BS$^m$, the instantaneous SINR is
\begin{equation}
	\gamma _{m,l,k}^{\left( u \right)} = \frac{{p_{m,l,k}^dg_{m,l,k}^{\left( u \right)}}}{{I_{m,l}^{{\rm{inter}},\left( u \right)}\left( {\bf{X}} \right) + I_{m,l}^{{\rm{intra}},\left( u \right)}\left( {\bf{X}} \right) + {N_0}}},
	\label{e14}
\end{equation}
where $p_{m,l,k}^d$ denotes the power for legitimate data, the channel gain $g_{m,l,k}^{\left( u \right)} = {\left| {{h_{{\rm{tu}}}}\left( {{f_j}} \right)} \right|^2}$ between BS$^m$ and UE follows an exponential distribution with parameter $\omega_u$. Therefore, the RTP of a terrestrial UE is given by
\begin{equation}
		{P_{t,u}}\!\left( {{\bf{X}},{\bf{p}}} \right) \!=\! \Pr\! \left( {\gamma _{m,l,k}^{\left( u \right)}  \!>\! {\tau _u}} \right) \!=\! \exp \left( { - \frac{{{N_0}{\tau _u}}}{{{\omega _u}p_{m,l,k}^d}}} \right)\!.
	\label{e15}
\end{equation}

In summary, these metric are used as a hard constraint in the optimization problem to specify the baseline reliability requirement for each active communication link.

\subsection{Problem Formulation} 
Cognitive secure transmission in STNs can be described as a constrained optimization problem, in which each time slot is driven and constrained by the environmental situation based on the real-time spectrum sensing result for the joint optimization of resources and security control. By applying link scheduling, we can design a secure and reliable decision-making framework, meeting high service demands and maximizing security. The matrix $\bf{S}$ selects the operating bands of all SAT, the matrix $\bf{X}$ specifies the frequency slots assigned by each controller to its UEs for transmission, the matrix $\bf{A}$ determines the locations of the adversarial signals to protect against eavesdropping, and the matrix $\bf{p}$ gives the legitimate and adversarial transmit powers over time-frequency positions. Except that $\bf{S}$ is a one-period global decision made by CN, the variables $\bf{X}$, $\bf{A}$, and $\bf{p}$ are jointly decided by each node with environment awareness (one node maps to a set of matrices). Accordingly, the overall problem can be formulated as
\begin{subequations}\label{e16}
	\setlength{\jot}{1.5pt}
	\renewcommand{\theequation}{\theparentequation{\alph{equation}}}
	\begin{align}
		\mathop {\max }\limits_{{\bf{S}},{\bf{X}},{\bf{A}},{\bf{p}}} \;\;&\sum\nolimits_z {\sum\nolimits_k {{P_e}\left( {{\bf{S}},{\bf{X}},{\bf{A}},{\bf{p}}} \right)} } \label{e16a}\\
		\text{s.t.}\quad\;&{P_e}\left( {{\bf{S}},{\bf{X}},{\bf{A}},{\bf{p}}} \right) \ge 1 - {\varepsilon _e},\forall z,k \label{e16b}\\
		&{P_u}\left( {{\bf{S}},{\bf{X}},{\bf{A}},{\bf{p}}} \right) \ge 1 - {\varepsilon _u},\forall z,k \label{e16c}\\
		&\sum\nolimits_n {{S_{n,j}}}  \le 1,\forall n,j \label{e16d}\\
		&\sum\nolimits_f {{X_{z,l,k,f}}}  = 1,\sum\nolimits_k {{X_{z,l,k,f}}}  \le 1,\label{e16e}\\ &\sum\nolimits_l {{X_{z,l,k,f}}}  \le 1,\forall z,l,k,f \label{e16f}\\
		&\sum\nolimits_f {{A_{z,l,k,f}}}  \le 1,\forall z,l,k,f \label{e16g}\\
		&\sum\nolimits_k {\left( {p_{z,l,k}^d + p_{z,l,k}^a} \right)}  \le p_z^{\max },\forall z,l,k \label{e16h}\\
		&{X_{z,l,k,f}} \le {S_{i,j}},\;{\rm{if}} \; z \in \mathcal{N_S},\forall l,k,f\label{e16i}
	\end{align}
\end{subequations}

In (15a), we maximize security performance, \textit{i.e.}, the sum of SP for all UEs. Constraints (15b) and (15c) enforce hard bounds on security and reliability, where ${\varepsilon _e}$ and ${\varepsilon _u}$ are small so that the decoding success probability for UE is close to one. Constraint (15d) restricts each band (multiple frequency slots) to a maximum of one SAT. Constraints (15e) and (15f) enforce a unique frequency slot for transmission and non-overlapping service per node. Constraint (15g) specifies the placement of the adversarial signal. Constraint (15h) states the node power budget for real and adversarial powers. Constraint (15i) couples $\bf{S}$ and $\bf{X}$, meaning a satellite UE can only be scheduled on frequency slot $f_j$ if its SAT has activated $f_j$.

Clearly, the optimization is both a non-convex combinatorial problem and an NP-hard problem, so existing exact methods are impractical. Moreover, since the CN and nodes must coordinate coupled variables such as time-frequency occupation and power, the setting with time-varying channels and partial information resembles a multi-agent stochastic game. In this context, MADRL can provide an approximate dynamic programming solution, enabling the learning of near-optimal policies through interaction with the environment and estimation of long-term returns. In addition, given the progress in generative AI, GANs can be employed to generate adversarial signals that increase detection error rates for eavesdroppers.

\section{Methods Design}
In this work, we decompose the decision-making process described by Eq.~(\ref{e16}) into three coupled stages. First, MADRL is employed to determine both $\bf{S}$ and $\bf{X}$. Second, GANs are adopted to decide $\bf{A}$ for adversarial insertion. Finally, we use MADRL to optimize $\bf{p}$ so as to attain the near-optimal security performance under reliability constraints.

\textbf{Stage I: MADRL for Frequency Slot Occupation:} To suppress interference at the macro level and ensure secure delivery at the micro level, we consider the joint design of the satellite operation matrix $\mathbf{S}$ and frequency slot occupation matrix $\mathbf{X}$. The process is defined by the tuple ${\left\langle {{\cal S},{\cal A},{\cal P},r,\gamma } \right\rangle }$, where $\mathcal{S}$ is the state space, $\mathcal{A}$ is the action space, $\mathcal{P}$ is the state transition probability, $r$ is the immediate reward, and $\gamma \in (0,1)$ is the discount factor. We adopt the centralized training and decentralized execution (CTDE) paradigm, \textit{i.e.}, global information is used during training while local observations are used for decision-making at execution for scalability and robustness\cite{b13}. The details are described as follows.

\emph{Agent:} The set of all agents is given by
\begin{equation}
	{\cal I} = \left\{ {{\rm{CN}}} \right\} \cup \left\{ {{\rm{SA}}{{\rm{T}}^{\rm{n}}}} \right\}_{n = 1}^N \cup \left\{ {{\rm{B}}{{\rm{S}}^{\rm{m}}}} \right\}_{m = 1}^M.
		\label{e17}
	\end{equation}
${{\rm{CN}}}$ outputs the global band plan $\bf{S}$, and each SAT$^n$ and BS$^m$ assigns frequency slots $\bf{X}$ to its UEs. At time $t \in \mathcal{L}$, agent $i \in {\cal I}$ receives a local observation and performs an action.

\emph{State:} The global state stacks: \textit{i)} the 0-1 indicators for active satellite frequency slots, which are from ${\bf{S}}^{N\times q}$; \textit{ii)} a discrete satellite coupling tensor $\left\{ {{{\bf{C}}^{\left( j \right)}}} \right\}_{j = 1}^q,{{\bf{C}}^{\left( j \right)}} \in \mathbb{R}_ + ^{N \times N}$; \textit{iii)} a binary vector of frequency slot occupation ${{\bf{s}}_{\bf{f}}}^{1 \times q}$ (0: idle, 1:busy); and \textit{iv)} a binary vector of user attachment ${{\bf{s}}_{\bf{u}}}^{1 \times K}$ (0:disconnected, 1:connected). Under CTDE, each agent has a local observation ${{o}}_i^{\left( t \right)} = {\Omega _i}\left( {{s^{\left( t \right)}}} \right) \in {{\cal{O}}_i}$. For CN, we have
\begin{equation}
o_{{\rm{CN}}}^{\left( t \right)} = \left(\! {{\rm{vec}}\left( {{{\bf{S}}^{\left( {t - 1} \right)}}} \right)\!,{\rm{vec}}\left( {{{\bf{C}}^{\left( 1 \right)}}} \right)\!, \ldots ,{\rm{vec}}\left( {{{\bf{C}}^{\left( {{q}} \right)}}} \right)} \!\right).
		\label{e18}
\end{equation}
For satellite SAT$^n$, the observation can be expressed as
\begin{equation}
	o_n^{\left( t \right)} = \left( {q_n^{\left( t \right)},{\bf{s}}_{{\bf{f}}|{q_n}}^{\left( t \right)},{{\cal{K}}_n}} \right),
	\label{e19}
\end{equation}
where ${q_n^{\left( t \right)}}$ denotes the satellite operating band index selected by ${{{\bf{S}}^{\left( {t} \right)}}}$, ${{\bf{s}}_{{\bf{f}}|{q_n}}^{\left( t \right)}}$ represents the local frequency slot occupation within that band, and ${{\cal{K}}_n}$ is the set of UEs served by SAT$^n$. For ground BS$^m$, the observation can be presented by
\begin{equation}
	o_m^{\left( t \right)} = \left( {\bf{s}}_{{\bf{f}}|{m}}^{\left( t \right)},{{\cal{K}}_m} \right).
	\label{e20}
\end{equation}
These observations can be updated in real time by spectrum sensing methods, such as recurrent neural networks and support vector machines (SVM)-based methods\cite{b34,b35}.

\emph{Action:} At each time step, the action of CN is
\begin{equation}
	a_{{\rm{CN}}}^{\left( t \right)} \!\equiv \!{{\bf{S}}^{\left( t \right)}} \!\in\! {{\cal A}_{{\rm{CN}}}}\!=\!\left\{ {{\bf{S}} \!\in\! {{\left\{ {0,1} \right\}}^{\!N \!\times q}}|\!\sum\nolimits_{\!n} \!{{S_{n,j}} \!\le \!1,\forall n} } \right\}\!,
	\label{e21}
\end{equation}
\textit{i.e.}, each satellite selects its operating band (a contiguous set of frequency slots) from among $q$ candidate frequency slots. The SAT$^n$ and BS$^m$ agents assign specific frequency slots to their respective UEs under a given $\bf{S}$, that is
\begin{equation}
	a_i^{\left( t \right)} \equiv {{\bf{X}}^{\left( t \right)}} \in {{\cal A}_i} = \prod\nolimits_{k \in {K_i}} {{{\cal{F}}_{i,k}}} ,
	\label{e22}
\end{equation}
where ${{{\cal{F}}_{i,k}}}$ is the set of available frequency slots for UE $k$.

\emph{Reward Function:} The objective is to maximize long-term security utility while satisfying stringent reliability constraints. Therefore, we use a global instantaneous reward as follows
\begin{equation}
	{r} = c_1 P_e + c_2 P_u - {c_3}{R_{\rm{occ}}} - {c_4}{R_{\rm{coll}}} - {c_5}{R_{\rm{tier}}}.
	\label{e23}
\end{equation}
Here, the primary considerations are the security metrics $P_e$ and reliability metrics $P_u$, which can be derived by scanning all frequency slots and combining the results with expressions. If the security requirement (${P_e} \ge 1 - {\varepsilon _e}$) is met, $c_1\!=\!1$; otherwise, it is set to zero. Similarly, only if ${P_u} \ge 1 - {\varepsilon _u}$ is met, $c_2\!=\!1$. ${R_{\rm{occ}}}$ measures the occupation cost and ${R_{\rm{occ}}}\!=\!1$ if the selected frequency slot is already occupied, which can be obtained via a row-by-row AND operation between ${{\bf{s}}_{\bf{f}}}$ and ${{\bf{X}}}$. ${R_{\rm{coll}}}$ penalizes conflicts within the same node and ${R_{\rm{coll}}}\!=\!1$ if multiple UEs occupy the same frequency slot at the same time, which can be determined by ${{\bf{X}}}$ within the current decision time. This term ensures that multiple UEs controlled by the same node do not experience mutual interference. ${R_{\rm{tier}}}$ penalizes inter- interference and ${R_{\rm{tier}}}=1$ if two nodes work at the same frequency slots, which can be obtained through the weighted sum of ${{\bf{S}}}$, ${{\bf{X}}}$ and ${{\bf{C}}}$ across all frequency slots. The parameters $c_3$, $c_4$, and $c_5$ are positive penalty weights.

We model the problem as a collaborative multi-agent sequential decision-making task in which all agents share a single global reward $r$ and maximize the team discount return
\begin{equation}
\mathop {\max }\limits_\pi  J\left( \pi  \right) = {\mathbb{E}_{\pi ,\mathcal{P}}}\left[ {\sum\limits_{t = t_1}^{t_L}  {{\gamma ^t}r\left( {{s^{\left( t \right)}},{a^{\left( t \right)}}} \right)} } \right],
	\label{e24}
\end{equation}
where $\gamma^t  \in \left( {0,1} \right)$, ${a^{\left( t \right)}} = {\bigl( {a_i^{\left( t \right)}} \bigr)_{i \in \mathcal{I}}}$ denotes joint actions. During the execution phase, a decentralized strategy is employed
\begin{equation}
	\pi \left( {a|o} \right) = \prod\nolimits_{i \in \mathcal{I}} {{\pi _i}\left( {{a_i}|{o_i}} \right)},
	\label{e25}
\end{equation}
where each agent $i$ independently performs actions based on its local observation ${o_i} = {\Omega _i}\left( s \right)$ for distributed collaboration. During the training phase, each agent learns a local action value ${Q_i}\left( {{o_i},{a_i};{\theta _i}} \right)$. A QMIX monotonic mixing network then aggregates these values to produce the team value as follows
\begin{equation}
	{Q_{{\rm{tot}}}}\!\left( {s,a;\Theta ,\psi } \right) \!=\! {f_{{\rm{mix}}}}\left( {{Q_1}, \ldots ,{Q_{\mathcal{I}}},s;\psi } \right)\!,\!\frac{{\partial {Q_{{\rm{tot}}}}}}{{\partial {Q_i}}} \ge 0,
	\label{e26}
\end{equation}
where $\Theta  = \left\{ {{\theta _i}} \right\}$, $\psi$ is the network parameter. As a result, this monotonicity can guarantee that
\begin{equation}
	\arg \mathop {\max }\limits_a {Q_{{\rm{tot}}}}\left( {s,a} \right) \Leftrightarrow {a_i} = \arg \mathop {\max }\limits_{{a_i}} {Q_i}\left( {{o_i},{a_i}} \right).
	\label{e27}
\end{equation}
Thus, execution can approach the optimal overall result without requiring centralized coordination. Within the CTDE framework, we employ DDQN to train each agent, with training samples written into the replay buffer as joint transition fragments $\left( {s,a,r,s'} \right)$. Specifically, the online network selects the joint action for the next state, which can be determined by
\begin{equation}
	a^*\left( {s'} \right) = \arg \mathop {\max }\limits_{a'} {Q_{{\rm{tot}}}}\left( {s',a';\Theta ,\psi } \right).
	\label{e28}
\end{equation}
While the target network evaluates the objective as follows
\begin{equation}
	y = r + \gamma {Q_{{\rm{tot}}}}\left( {s',a^*\left( {s'} \right);{\Theta ^ - },{\psi ^ - }} \right),
	\label{e29}
\end{equation}
where ${\Theta }$, ${\Theta ^ - }$ and ${\psi}$, ${\psi ^ - }$ denote the parameters of the online network and target network, respectively. Then minimize
\begin{equation}
	{\cal{L_S}} = \mathbb{E}\left[ {{{\left( {y - {Q_{{\rm{tot}}}}\left( {s,a;\Theta ,\psi } \right)} \right)}^2}} \right].
	\label{e30}
\end{equation}
During execution, each agent uses the $\varepsilon$-greedy method within the set of feasible actions defined by the constraints.

\begin{figure*}[t!]
	\centerline{\includegraphics[width=1\linewidth]{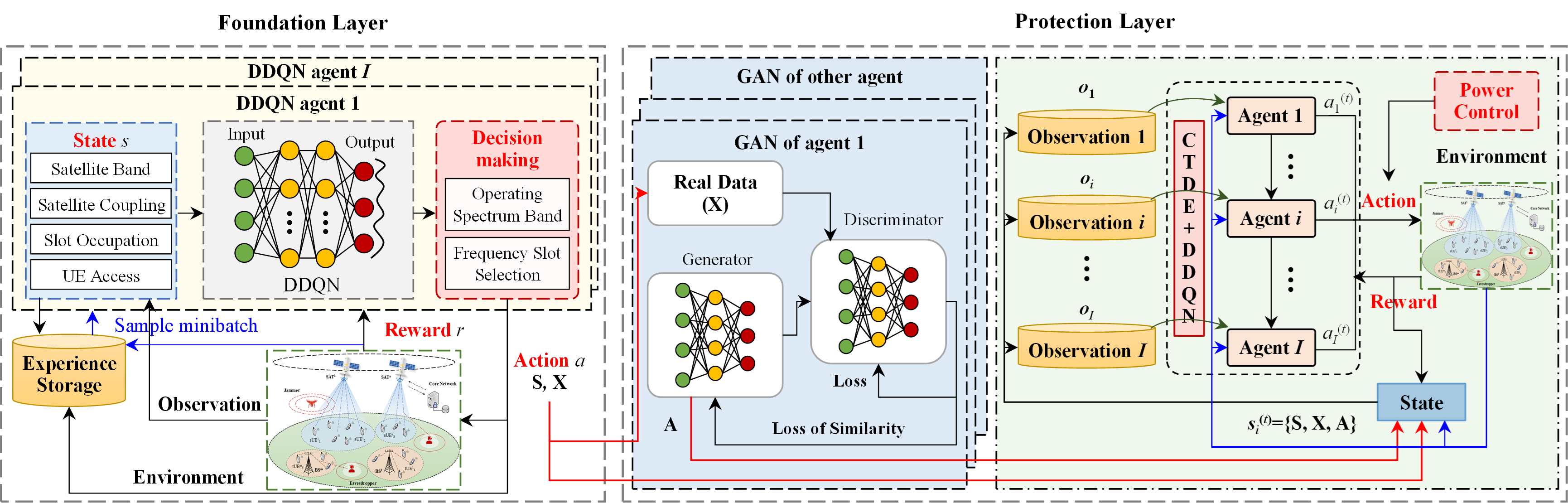}}
	\setlength{\belowcaptionskip}{-0.5cm}   
	\caption{Illustration of the proposed multi-agent-driven decision-making framework. Our method builds a two-layer defense system. On the left, the foundation layer employs MADRL to decide the matrix $\bf{S}$ and $\bf{X}$. The protection layer has two parts. First, given $\bf{S}$ and $\bf{X}$, GANs are trained to produce an adversarial matrix $\bf{A}$. Second, with $\bf{S}$, $\bf{X}$ and $\bf{A}$ fixed, we determine the optimal power within the feasible region. The red arrows can illustrate the flow of training data across modules. }
	\label{f4}
\end{figure*}

\textbf{Stage II: GANs for Adversarial Insertion:} GANs forms a zero-sum game between a generator $G$ and a discriminator $D$. The generator network generates synthetic data samples from the training set designed to mimic real data. The discriminator network then evaluates these samples, distinguishing between real and fake data, and aiming to correctly identify whether a given sample is genuine or fabricated. Hence, this competition encourages both networks to improve their performance repeatedly, making GANs as a useful tool for us to generate adversarial signals in the wireless system\cite{b2,b14,b26}.

In our design, the adversarial patterns are required to match the learned matrix ${{\bf{X}}}$ in terms of structure and statistical behavior, so that the discriminator is confused and thus misjudges the data. Specifically, the generator $\!G\!$ takes random noise $\!z \!\sim\! p_z(z)$ as input and produces a synthetic spectrum allocation $\!G(z)\!$. The discriminator $\!D\!$ receives both the real matrix $x \!\!\sim\! p_{data}(x)$ and the generated matrix $G(z)$, learning to classify them and outputting the judgment result. For clarity, the training procedure of $G$ and $D$ can be achieved through the minimization and maximization objective function 
\begin{equation}
	\small
	\mathop {\min }\limits_G \!\mathop {\max }\limits_D \!{\mathbb{E}_{x \sim\! {p_{d\!a\!t\!a}}\!\left( \!x \right)}}\!\left[ {\log \!D\left( x \right)} \right] \!+\! {\mathbb{E}_{z \sim\! {p_z}\left( \!z\! \right)}}\!\left[ {\log \big( {1 \!-\! D\!\left( {G\!\left( z \right)} \right)} \big)} \!\right]\!.
	\label{e31}
\end{equation}

In fact, the discriminator aims to force $D\left( x \right)$ towards 1 (correctly identifying the legitimate matrix) and $D\!\left( {G\!\left( z \right)}\right)$ towards 0 (detecting the generated matrix). Conversely, the generator seeks to increase $D\!\left( {G\!\left( z \right)}\right)$ values (deceiving the discriminator into accepting the generated matrix). We adopt the Wasserstein GAN with a gradient penalty (WGAN-GP) to improve stability. During training, the generator optimizes its parameters to make the generated adversarial matrix ${\bf{A}}$ resemble the real occupation matrix ${\bf{X}}$ in terms of their spectral characteristics and energy distribution. The discriminator acts as a critic with a real-valued output, and its loss is defined by
\begin{equation}
	{\cal{L_D}} = {\mathbb{E}_{\bf{A}}}\left[ {D\left( {\bf{A}} \right)} \right] - {\mathbb{E}_{\bf{X}}}\left[ {D\left( {\bf{X}} \right)} \right] + {\lambda _{\rm{GP}}}{\cal{L_{\rm{GP}}}},
	\label{e32}
\end{equation}
where ${\cal{L_{\rm{GP}}}} = {\mathbb{E}_{{\bf{\hat X}}}}\bigl[ {{{\bigl( {{{\bigl\| {{\nabla _{{\bf{\hat X}}}}D\bigl( {{\bf{\hat X}}} \bigr)} \bigr\|}_2} - 1} \bigr)}^2}} \bigr]$ and ${\bf{\hat X}} = \rho {\bf{X}} + \left( {1 - \rho } \right){\bf{A}}$ with $\rho  \sim {\cal{U}} \left( {0,1} \right)$. The first two terms enforce the Wasserstein margin between fake and real data, and the third term is the gradient penalty. In addition, the generator loss can be designed as follows
\begin{equation}
	{\cal{L_G}} =  - {\mathbb{E}_{\bf{A}}}\left[ {D\left( {\bf{A}} \right)} \right] + \alpha  \cdot {{\cal{L}}_{{\rm{sim}}}} + \beta  \cdot {{\cal{L}}_{{\rm{occ}}}},
	\label{e33}
\end{equation}
with the loss function
\begin{equation}
 {{\cal{L}}_{{\rm{sim}}}} = {\left\| {{\bf{A}} - {\bf{X}}} \right\|_1} + {D_{{\rm{KL}}}}\left( {{P_{\bf{A}}}||{P_{\bf{X}}}} \right),
	\label{e34}
\end{equation}
to match support and structure, and
\begin{equation}
	\small
	{{\cal{L}}_{{\rm{occ}}}} \!=\!\! {\sum\limits_{l = 1}^L \!\!{\sum\limits_{f = {f_1}}^{{f_q}}\!\! {\left(\! {\frac{1}{K}\!\left( {\sum\limits_{k = 1}^K {{\bf{1}}\!\left( {{A_{z,l,k,f}} \!=\! 1} \right) \!-\!\! \sum\limits_{k = 1}^K \!{{\bf{1}}\!\left( {{X_{z,l,k,f}} \!=\! 1} \right)\!} } } \right)}\! \right)} } ^2},
	\label{e35}
\end{equation}
to keep load statistics within a controlled range\cite{b36}.

\textbf{Stage III: MADRL for Power Control:} Since the system has fixed the frequency slot occupation matrix ${{\bf{X}}}$ for legitimate traffic and the adversarial insertion matrix ${{\bf{A}}}$ to increase the difficulty of eavesdropper detection, power control is required for both signals at each time-frequency position. To obtain the optimal powers given $\left(\bf{S},\bf{X},\bf{A}\right)$, we train all agents with DDQN to learn the power matrix $\mathbf{p}=\bigl\{ {{p^d_{z,l,k,f}}},{{p^a_{z,l,k,f}}} \bigr\}$. The relevant decision elements are defined as follows.

\emph{Agent:} At this point, the set of agents comprises all network nodes $z$ (\textit{i.e.}, SATs and BSs). Having already received their local frequency slot assignments and adversarial locations, they now need to decide on legitimate and adversarial powers.

\emph{State:} The information observed by each agent includes: \textit{i)} the current frequency slot occupation of all UEs under its control, uniquely determined by matrix ${{\bf{X}}}$; \textit{ii)} the deployment status of adversarial signals at the current decision moment, uniquely determined by matrix ${{\bf{A}}}$; \textit{iii)} the available power of each agent with peak and possibly average limits.

\emph{Action:} At each decision step, each agent selects the power levels $\bigl\{ {{p^d_{z,l,k,f}}},{{p^a_{z,l,k,f}}} \bigr\}$ for transmitting legitimate and adversarial signals, and these levels must within the feasible set.

\emph{Reward Function:} As the security and reliability of information both depend on power, we take the SP as positive feedback. Although adversarial signals are deployed on the time-frequency channel, they carry no meaningful information. Transmitting these signals at high power levels results in significant power consumption. Therefore, the instantaneous reward is designed as follows
\begin{equation}
	{r}' = d_1 P_e  - {d_2}{{{K_z}p_{z,l,k}^a} \mathord{\left/
			{\vphantom {{{K_z}p_{z,l,k}^a} {p_z^{\max }}}} \right.
			\kern-\nulldelimiterspace} {p_z^{\max }}},
	\label{e36}
\end{equation}
where $p_{z,l,k}^a$ is the adversarial power used by node $z$ in the current time slot, ${p_z^{\max }}$ is the total transmit power of node $z$, and $K_z$ is the total number of legitimate UEs associated with the node $z$. The weights $d_1$ and $d_2$ balance the impact of resource competition and power cost, where $d_1$ has the same definition as $c_1$ and $d_2$ is a constant.

\textbf{Algorithm Design:} The offline training stage of our method is shown in Algorithm~\ref{alg1}. In our framework, these three stages are sequentially trained offline. First, MADRL is applied under CTDE to train all agents until convergence. This step learns the optimal satellite band operation matrix ${{\bf{S}}}$ and frequency slot occupation matrix ${{\bf{X}}}$, aiming to maximize security while preserving reliability. Next, we use the matrix ${{\bf{X}}}$ as a structural anchor to train GANs, which synthesizes an adversarial matrix distribution ${{\bf{A}}}$. Shape similarity and slot sparsity are enforced to control the pattern without affecting traffic delivery. Finally, with the positions of ${{\bf{A}}}$ fixed, we employ multi-agent learning to control power, allowing us to obtain the power matrix ${{\bf{p}}}$. All three stages adopt experience replay and target networks for stable convergence. The online inference of our method is shown in Algorithm~\ref{alg2}. Training is not performed during online decision-making. Instead, forward inference is performed in a single step at each decision step by following this sequence. A slow timescale at the core plans operating bands and constrains the action space, whereas a fast timescale lets network nodes execute the learned policies in a decentralized manner.

\begin{algorithm}[!t]
	\caption{Offline Training Stages of the Proposed Multi-Agent-Driven Transmission Method}
	\label{alg1}
	\renewcommand{\algorithmicrequire}{\textbf{Input:}} 
	\renewcommand{\algorithmicensure}{\textbf{Output:}}
	\begin{algorithmic}[1]
		\Require Agent set ${\cal I}$ (CN, SAT, BS) as in Eq.~(\ref{e17}); candidate frequency slots $q$; coupling $\{\mathbf{C}^{(j)}\}_{j=1}^q$; discount $\gamma$; security target $1-\varepsilon_e$; reliability target $1-\varepsilon_u$; DDQN hyperparameters $(\eta, T_{\text{target}})$; WGAN-GP hyperparameters $(\lambda_{\rm GP},\alpha,\beta,n_{\rm critic})$
		\Ensure Trained policies for $(\mathbf{S},\mathbf{X})$, generator $G$, and power policy yielding $(\mathbf{S},\mathbf{X},\mathbf{A},\mathbf{p})$
		\State \textbf{Init} Stage-I per-agent $Q_i$ \& mixer $f_{\rm mix}$; Stage-III $Q_i$ \& mixer; WGAN critic $D$, generator $G$; buffers $\mathcal{D},\mathcal{D}'$
		\Statex \textbf{\textit{Stage I: MADRL for $\mathbf{S},\mathbf{X}$ (CTDE)}}
		\For{ep $=1{:}E_1$}
		\State Reset env; $s\!\gets$ init; $o_i\!\gets\!\Omega_i(s)$ per Eqs.~(\ref{e18})--(\ref{e20}) 
		\For{$t=t_1{:}t_L$}
		\State CN samples $a_{\rm CN}^{(t)}\!\equiv\!\mathbf{S}^{(t)}\in \mathcal{A}_{\rm CN}$ via $\varepsilon$-greedy
		\State Nodes sample $a_i^{(t)}\!\equiv\!\mathbf{X}^{(t)}\in \mathcal{A}_i$ via $\varepsilon$-greedy
		\State Execute $a^{(t)}$; compute global $r$ by Eq.~(\ref{e23}); observe $s'$; store $(s,a,r,s')$ in $\mathcal{D}$
		\State \textsc{UpdateQmix}($\mathcal{D}$; use Eqs.~(\ref{e28})--(\ref{e30}) and monotonicity Eqs.~(\ref{e26})--(\ref{e27})); target sync every $T_{\text{target}}$
		\State $s\!\gets\!s'$, $o_i\!\gets\!\Omega_i(s)$
		\EndFor
		\EndFor
		\State Decode $\mathbf{S},\mathbf{X}$ via greedy $\{\pi_i\}$ with factorization Eq.~(\ref{e25})
		\Statex \textbf{\textit{Stage II: WGAN-GP for $\mathbf{A}$ (match $\mathbf{X}$)}}
		\For{iter $=1{:}E_2$} 
		\State \textsc{UpdateWGAN-GP}: run $n_{\rm critic}$ critic steps with loss Eq.~(\ref{e32}); then one generator step with Eqs.~(\ref{e33})--(\ref{e35})
		\EndFor
		\State $\mathbf{A}\!\gets\!G(z)$
		\Statex \textbf{\textit{Stage III: MADRL for Power $\mathbf{p}$ (fixed $(\mathbf{S},\mathbf{X},\mathbf{A})$)}}
		\For{ep $=1{:}E_3$}
		\State Reset env with fixed $(\mathbf{S},\mathbf{X},\mathbf{A})$; build power obs
		\For{$t=t_1{:}t_L$}
		\State Each node $z$ chooses $\{p^d_{z,l,k,f},p^a_{z,l,k,f}\}$ in feasible set (non-negativity and peak/avg)
		\State Execute power action; compute $r'$ by Eq.~(\ref{e36}); store $(s,a,r',s')$ in $\mathcal{D}'$
		\State \textsc{UpdateQmix}$(\mathcal{D}';\ \text{same as Stage I})$
		\EndFor
		\EndFor
		\State Obtain $\mathbf{p}$ via greedy decoding of trained policies
	\end{algorithmic}	
\end{algorithm}
\setlength{\belowcaptionskip}{-12pt}
\setlength{\textfloatsep}{12pt}

\begin{algorithm}[t]
	\caption{Online Inference Stages of the Proposed Method (Single-Step per Decision Point)}
	\label{alg2}
	\begin{algorithmic}[1]
		\Require Trained models from Stages I-III; current observations (spectrum sensing, attachments, occupation); timescale design (slow band planning and fast slot/power)
		\Ensure Actions $(\mathbf{S},\mathbf{X},\mathbf{A},\mathbf{p})$ at the current decision epoch
		\State \textbf{Slow timescale (core network):} The CN decodes $\mathbf{S}=\arg\max_{a_{\rm CN}} Q_{\rm tot}(s,a;\Theta,\psi)$ subject to Eq.~(\ref{e21})
		\State \textbf{Fast timescale (node):} SATs and BSs decode $\mathbf{X}$ with Eq.~(\ref{e22}); generator outputs $\mathbf{A}=G(z)$ aligned to $\mathbf{X}$; nodes output $\mathbf{p}$ adopting trained power policies
		\State Enforce feasibility masks (occupation, conflict, inter-interference, and intra-interference) and reliability constraint $(P_u \ge 1-\varepsilon_u)$ from Eqs.~(\ref{e23}) and (\ref{e36})
		\State Execute $(\mathbf{S},\mathbf{X},\mathbf{A},\mathbf{p})$; no parameter updates online
	\end{algorithmic}
\end{algorithm}

The diagram of our proposed method is shown in Fig.~\ref{f4}. Based on the above idea and framework, our design provides a two-layer defense system for STNs. The first layer is the \textbf{Foundation Layer}, in which we adopt MADRL to schedule links on both time and frequency resources in an interference-free and orderly manner. This structured scheduling makes detection more difficult for an eavesdropper and establishes a security baseline with reliable access. The second layer is the \textbf{Protection Layer}. Here, we employ GANs to derive an optimal adversarial signaling pattern whose time-frequency occupation and sparsity statistics align with the baseline scheduling. We then apply power control to legitimate and adversarial signals of each node, which can disrupt the decision of the eavesdropper with system constraints. Taken together, these two layers provide structured anti-eavesdropping capability while maintaining reliability and efficient resource usage.

\textbf{Computational Complexity Analysis:} Let there be $Z$ agents (CN, $N$ SAT, and $M$ BSs). Agent $i$ employs a value network with widths $\bigl\{ {B_{\ell ,i}^x} \bigr\}_{\ell  = 0}^{L_i^x}$. Denote the QMIX mixer parameter size by $\left| {{\theta _{{\rm{mix}}}}} \right|$ and its per-update cost by ${\cal O}\left( {\left| {{\theta _{{\rm{mix}}}}} \right|} \right)$. With mini-batch size ${\cal B}$, trajectory length $T_1$, and $I_1$ training iterations, one DDQN can cost ${\Xi _x}={\cal O}\bigl( {\sum\nolimits_{i = 1}^Z {\bigl( {B_{0,i}^xB_{1,i}^x + \sum\nolimits_{\ell  = 1}^{L_i^x - 1} {B_{\ell ,i}^xB_{\ell  + 1,i}^x} } \bigr) + \left| {{\theta _{{\rm{mix}}}}} \right|} } \bigr)$. Hence, the total offline cost of Stage-I is ${\cal O}\left( {{I_1}{\cal B}{T_1}{\Xi _x}} \right)$\cite{b37}. Let $\left| {{\theta _G}} \right|$ and $\left| {{\theta _D}} \right|$ be the parameter counts of $G$ and $D$, respectively. One generator-discriminator round includes $E_g$ critic updates ${\cal O}\left( {\left| {{\theta _D}} \right|} \right)$ with GP, and one generator update ${\cal O}\left( {\left| {{\theta _G}} \right|} \right)$. We denote their aggregated per-round cost by ${\Omega _{{\rm{sim/occ}}}}$. Over $I_g$ rounds, the total Stage-II cost is ${\cal O}\left( {{I_g}\left[ {{E_g}\left| {{\theta _D}} \right| + \left| {{\theta _G}} \right| + {\Omega _{{\rm{sim/occ}}}}} \right]} \right)$\cite{b38}. Let the set of power-control agents be all nodes. Agent $i$ uses widths $\bigl\{ {B_{\ell ,i}^p} \bigr\}_{\ell  = 0}^{L_i^p}$. With mini-batch ${\cal B}'$, trajectory length $T_3$, and $I_3$ iterations, one DDQN costs ${\Xi _p}= {\cal O}\bigl( {\sum\nolimits_{i = 1}^Z {\bigl( {B_{0,i}^pB_{1,i}^p + \sum\nolimits_{\ell  = 1}^{L_i^p - 1} {B_{\ell ,i}^pB_{\ell  + 1,i}^p} } \bigr) + \left| {{\theta _{{\rm{mix}}}}} \right|} } \bigr)$, and therefore the total Stage-III cost is ${\cal O}\left( {{I_3}{\cal B}'{T_3}{\Xi _p}} \right)$. Hence, the one-time training complexity (performed at ground stations) is ${\cal O}\left( {{I_1}B{T_1}{\Xi _x} + {I_g}\left[ {{E_g}\left| {{\theta _D}} \right| + \left| {{\theta _G}} \right| + {\Omega _{{\rm{sim/occ}}}}} \right] + {I_1}B{T_1}{\Xi _p}} \right)$.

At each decision epoch, the system executes: \textit{i)} one forward pass of the Stage-I policies for CN, SAT, and BS; \textit{ii)} one generator call $G\left(z\right)$ with pre-loaded weights to obtain $\mathbf{A}$; \textit{iii)} one forward pass of the Stage-III policies for power control; \textit{iv)} feasibility masking and light-weight lookups (\textit{e.g.}, conflict checks and reliability gating). Let ${{\tilde \Xi }_x}$ and ${{\tilde \Xi }_p}$ denote the forward-only costs of the trained networks. If at most $K'N' \le {10^3}$ candidate entries are validated per step, the complexity at each time slot is ${\cal O}\bigl( {{{\tilde \Xi }_x} + \left| {{\theta _G}} \right| + K'N'{{\tilde \Xi }_p}} \bigr)$\cite{b39}. In practice, with modest hidden widths, the above cost comfortably meets millisecond-level decision deadlines.

\section{Performance Evaluation}

This section provides simulation results to evaluate the proposed multi-agent-driven cognitive secure communication method. We consider a network covered by two LEO satellites and four BSs. The total bandwidth of 800 MHz is shared equally among all nodes and divided into 64 non-overlapping frequency slots, with each satellite permitted to occupy up to 32 contiguous frequency slots as its operating band. One transmission period consists of 64 time slots. The transmit power is set to 53 dBm for SATs and 37 dBm for BSs\cite{b40}. For legitimate UEs, the decoding threshold is 2 set to dB, the security constraint is fixed at 0.8, and the reliability constraint varies from 0.9 to 0.99. The number of UEs ranges from 4 to 64. For eavesdroppers, the decoding threshold is also set to 2 dB, and they are randomly located in the network\cite{b41}. We model three eavesdropping modes: energy detection, SVM-based detection, and DRL-predicted detection. The background noise power is -103 dBm. The channel parameters are set to $\Omega _u=5$ and $\Omega _e=3$, and $\omega _u=1$ and $\omega _e=0.75$. The reward weights are $c_3=c_4=c_5=0.1$, and $d_2=0.5$, with a learning rate of 0.05 and a discount factor of 0.98\cite{b42}.

We compare the proposed transmission method (denoted as MADRL-GANs) with the following methods:

\textit{i)} Artificial-noise (AN) assisted frequency hopping (denoted as AN-Assisted FH). A series of pseudo-random logistic-map sequences provides a hopping pattern with high randomness and a long period. AN is superposed on top of useful signals to reduce the correct detection probability of eavesdropper\cite{b43}.

\textit{ii)} Game-theoretic secure resource allocation (denoted as Game-Theoretic Allocation). The legitimate network and the eavesdropper are modeled as players with a commitment or observability advantage. The leader announces power or beam strategies, to which the follower reacts accordingly\cite{b44}.

\textit{iii)} GANs-based power control (denoted as GANs-Based Control). At the relay and forwarding nodes, GANs are trained so that the generator produces adaptive power allocation schemes for secure transmission, and the discriminator decides whether to release legitimate messages\cite{b45}. 

\begin{figure}[!t]
	\centering
	\vspace{-0.2cm}  
	\setlength{\abovecaptionskip}{-0.25cm}   
	\setlength{\belowcaptionskip}{-0.3cm}   
	\begin{minipage}{0.75\linewidth}
		\centering
		\includegraphics[width=\linewidth]{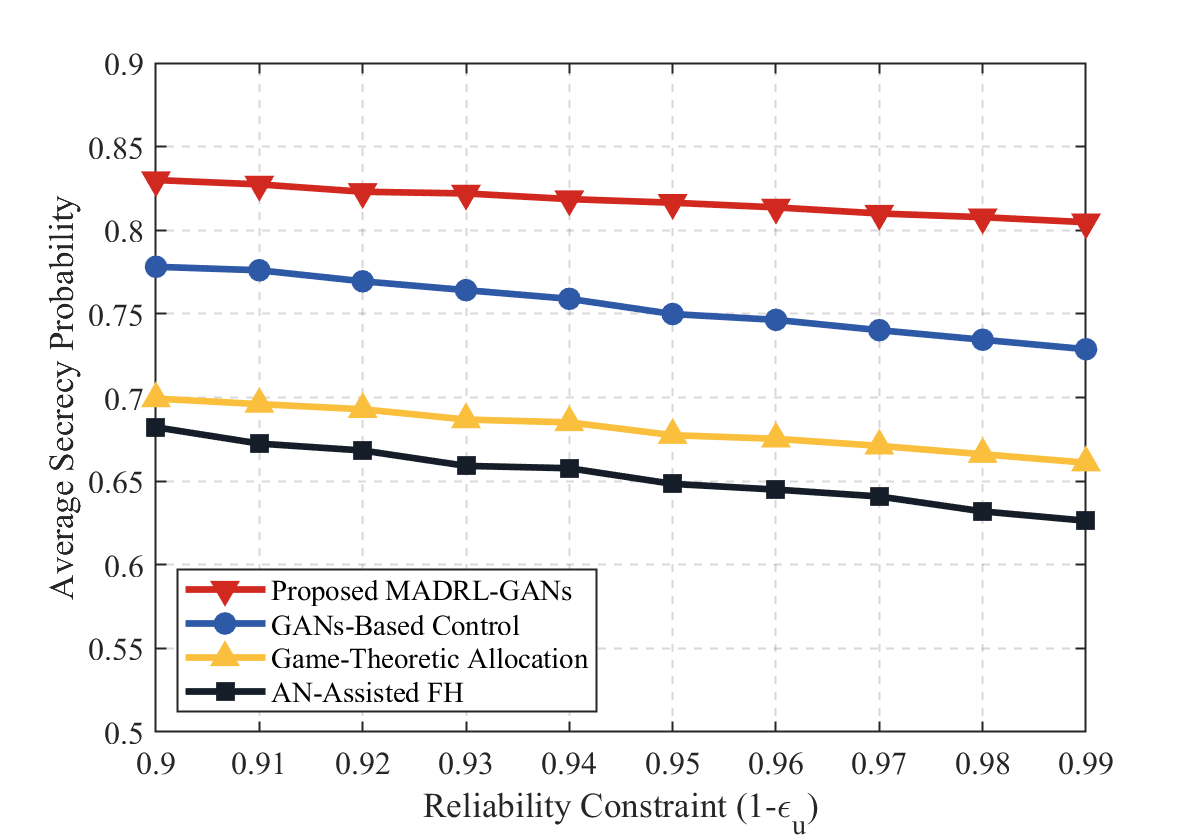}
		\parbox{\linewidth}{\centering (a)} 
		\label{f5a}
	\end{minipage}
	
	\vspace{-0.35cm}
	
	\begin{minipage}{0.75\linewidth}
		\centering
		\includegraphics[width=\linewidth]{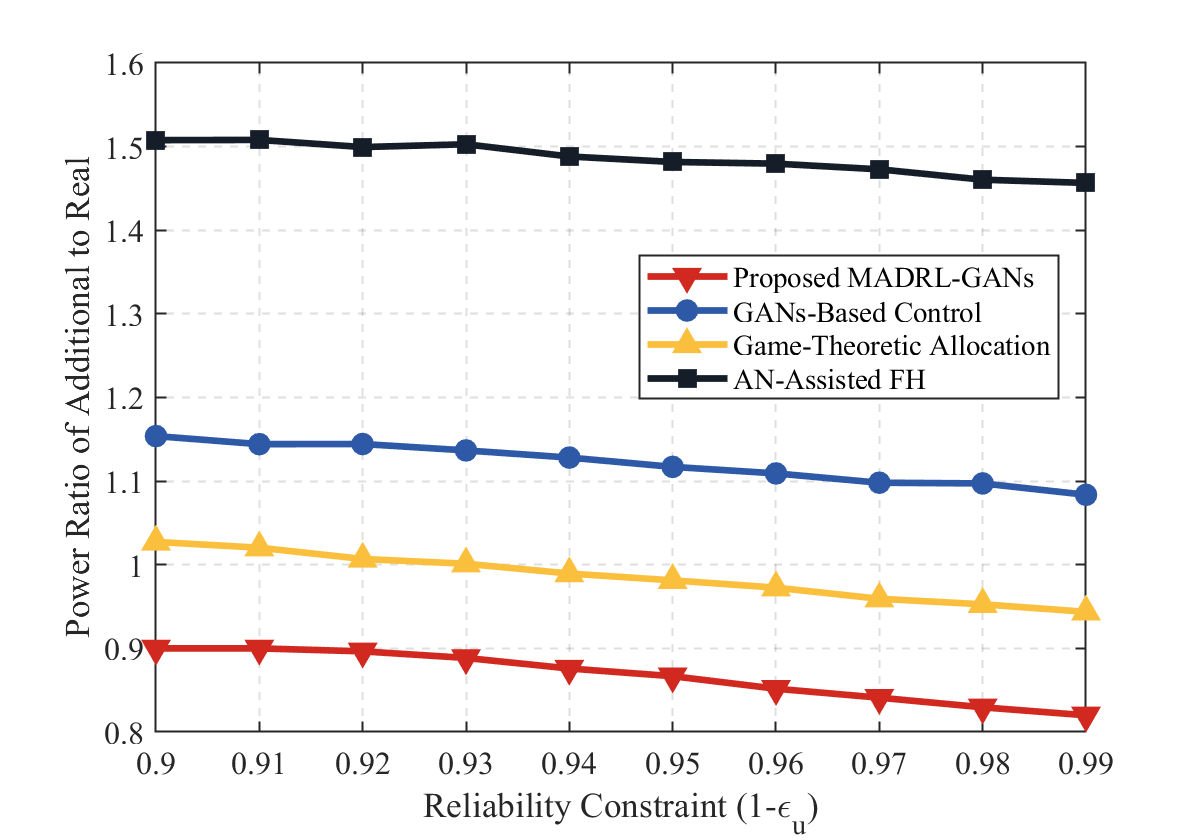}
		\parbox{\linewidth}{\centering (b)} 
		\label{f5b}
	\end{minipage}
	\caption{Performance comparisons vs. different reliability.}
	\label{f5}
\end{figure}

\begin{figure}[!t]
	\centering
	\vspace{-0.2cm}  
	\setlength{\abovecaptionskip}{-0.25cm}   
	\setlength{\belowcaptionskip}{-0.3cm}   
	\begin{minipage}{0.75\linewidth}
		\centering
		\includegraphics[width=\linewidth]{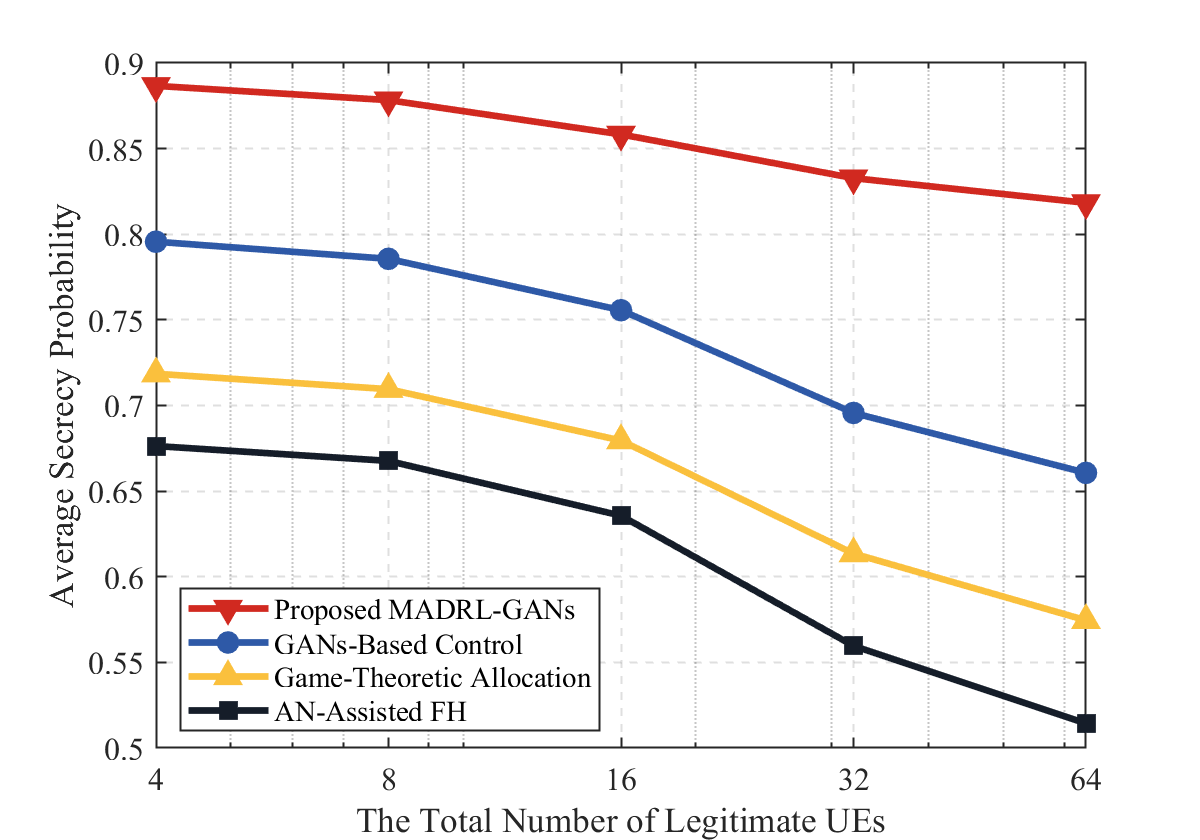}
		\parbox{\linewidth}{\centering (a)} 
		\label{f6a}
	\end{minipage}
	
	\vspace{-0.35cm}
	
	\begin{minipage}{0.75\linewidth}
		\centering
		\includegraphics[width=\linewidth]{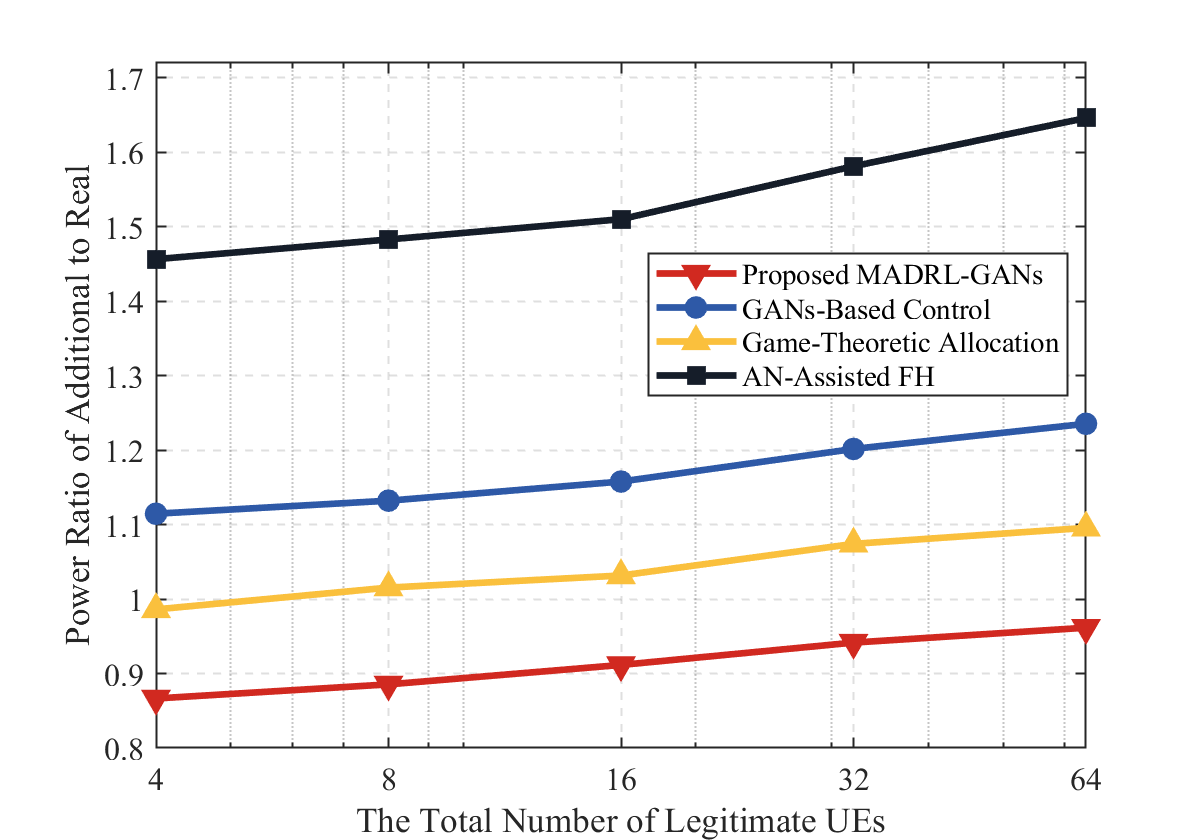}
		\parbox{\linewidth}{\centering (b)} 
		\label{f6b}
	\end{minipage}
	\caption{Performance comparisons vs. different UEs numbers.}
	\label{f6}
\end{figure}

\subsection{Performance Comparisons Under Different Reliability}
Fig.~\ref{f5} shows the performance of all methods under different reliability constraints when there are 16 legitimate UEs and the eavesdropper adopts DRL-predicted detection. As can be seen in Fig.~\ref{f5}, for all methods, the SP and the power ratio drop as the target reliability constraint increases. This is because stricter reliability constraints force the system to allocate more power to the legitimate signal within a limited power budget. This reduces the available adversarial injection power, and thus weakens security performance. In this case, both the transmit power and the frequency slot occupation scheme need to be jointly optimized to guarantee network performance.

We also observe from Fig.~\ref{f5}(a) that within a region of change in reliability constraints, the other two methods can still achieve a high SP compared with AN-assisted FH and game-theoretic allocation methods. However, if the reliability constraints are extremely strict (\textit{e.g.}, the reliability constraint grows beyond 0.98), there will be more unwanted information leakage events. Fig.~\ref{f5}(b) shows that the AN-assisted FH method generally requires a higher power cost to achieve a comparable level of security to the other three methods. In aggregate, our proposed method achieves a higher SP and lower power consumption under different reliability requirements, and the more stringent the reliability constraint, the greater the performance gap in its favor. The reason is that our method adopts MADRL to optimize the frequency slot occupation and power control, and employs GANs to generate time-frequency aligned adversarial signals that disrupt the DRL-predicted detection adopted by the eavesdropper. Notably, Fig.~\ref{f5}(a) also shows a representative trend that, in comparison with the other methods, the SP curve of our method degrades more slowly as the reliability constraint tightens. The reason is that each decision step fuses online spectrum sensing with penalties for inter- and intra- interference in the reward, which enables us to avoid harmful interference during transmission, thus meeting the reliability target while maintaining a high security level.

\subsection{Performance Comparisons Under Different UEs Numbers}
Fig.~\ref{f6} shows the performance comparisons for different secure methods with respect to increasing numbers of UEs, when the reliability constraint is 0.95 and the eavesdropper employs DRL-predicted detection. As the number of UEs increases within the admissible network range (from 4 to 16), we observe a slight drop in SP and a modest rise in power consumption for all methods. However, there is a sharp change when the number of UEs grows beyond the admissible range. A larger set of UEs intensifies spectrum contention and makes link scheduling and resource control more complex. With more active connections, the eavesdropper can identify more stable time-frequency occupation patterns, which increases the chance of hitting occupied frequency slots. Besides, a higher load results in stronger co-channel interference over a longer period. To satisfy the security command, nodes must allocate additional power, thereby increasing the power cost. As shown in Fig.~\ref{f6}, with a small number of UEs, SP declines slowly with load; however, under large-scale access, both security and power efficiency deteriorate more visibly.

From Fig.~\ref{f6}, although all methods experience a drop in SP as the number of UEs increases, the proposed MADRL-GANs method still achieves better security performance and lower power cost than the other three methods. Remarkably, the proposed method obtains a SP of over 85\% when the UE count is below 16, and the adversarial power is lower than the legitimate data power. When the number of UEs exceeds 16, the SP declines more markedly, yet this reduction is the slowest among all methods. In the meantime, the adversarial power cost becomes comparable to the legitimate signal power. This behavior can be explained by resource contention under limited spectrum and power budgets. As the load increases, meeting the reliability constraint requires more power to be allocated to the legitimate link, reducing the power available for adversarial injection and weakening its masking effect.

\begin{figure}[t!]
	\centerline{\includegraphics[width=0.75\linewidth]{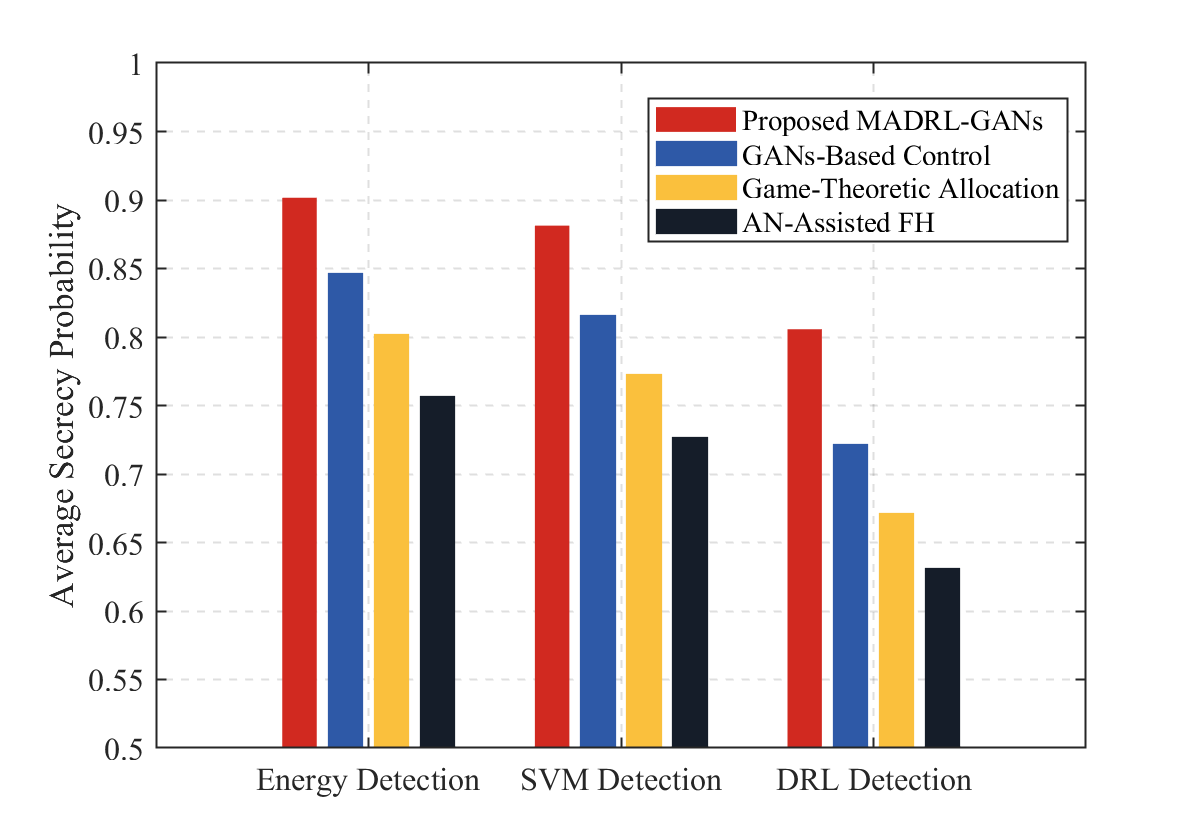}}
	\caption{Performance comparisons vs. eavesdropper models.}
	\label{f7}
	\vspace{0.2cm}
\end{figure}

\subsection{Performance Comparisons Under Different Eavesdroppers}
Fig.~\ref{f7} shows the performance comparisons under a reliability constraint of 0.95 with 16 UEs, considering different eavesdropper models. We consider three eavesdroppers. Firstly, an energy detection-based eavesdropper randomly selects several frequency slots and uses a radiometer to test for signal presence. Secondly, an SVM-based eavesdropper randomly monitors several frequency slots and applies a supervised classifier to the observed samples. Thirdly, a DRL-predicted eavesdropper acts as an agent that can learn time-frequency occupation from past periods, predict the next slot usage and perform focused detection to increase the hitting probability. From Fig.~\ref{f7}, for all methods, the SP is highest against energy detection and lowest against the DRL-based eavesdropper. This is because the DRL model provides policy learning and prediction, which enables more accurate targeting of likely occupied frequency slots, thereby posing a stronger threat.

From Fig.~\ref{f7}, we also find that our method achieves the highest SP against all three eavesdropper models. The improvement does not only come from the optimization of frequency slot occupation and power allocation under a security-oriented objective. A key factor is the use of adversarial signals that match the structure and sparsity of real traffic. This design increases the detection error of eavesdropper, thus weakening its ability to identify useful payloads. Even when the eavesdropper performs learning and prediction, statistical alignment of the adversarial signal induces misclassification of active frequency slots, reducing the chance of correct interception.

\begin{figure}[t!]
	\centerline{\includegraphics[width=0.75\linewidth]{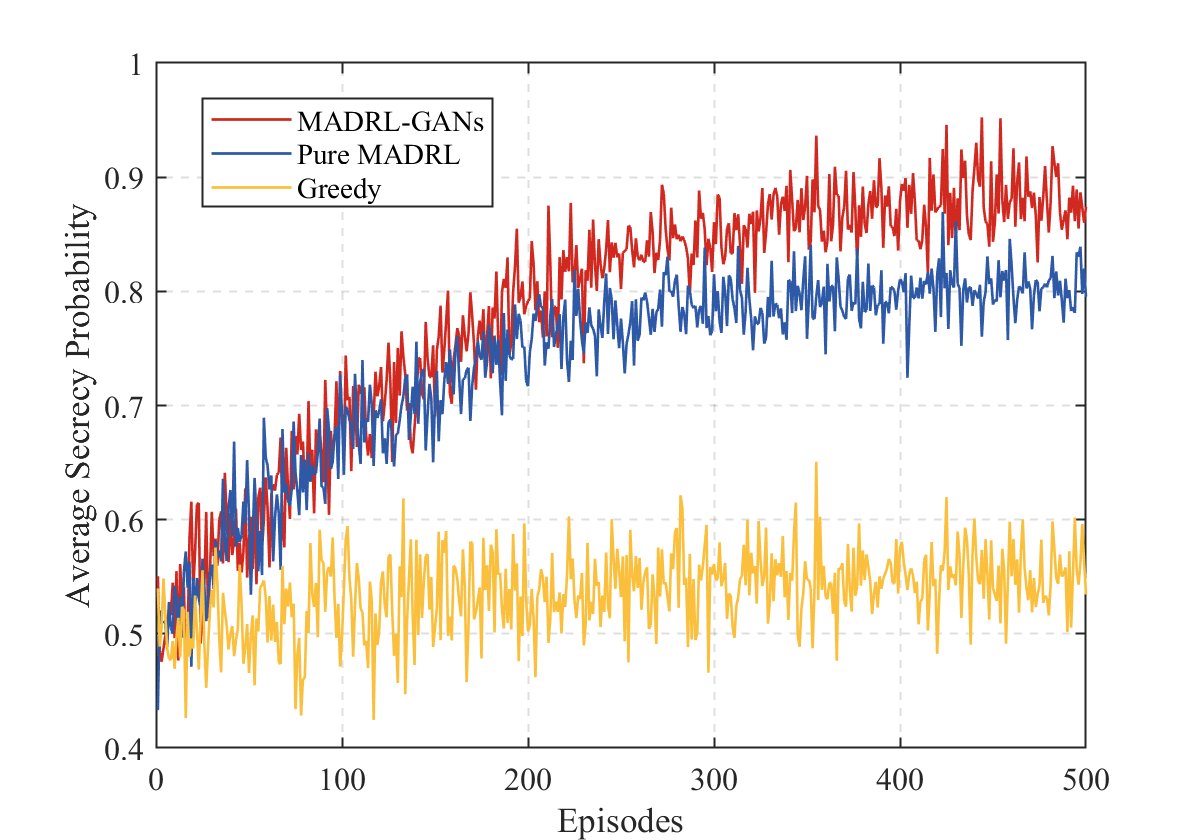}}
	\caption{Convergence comparisons of compared methods.}
	\label{f8}
	\vspace{0.2cm}
\end{figure}

\subsection{Convergence Comparisons}
We show in Fig.~\ref{f8} the SP with increasing training episodes to examine the convergence behavior of the proposed method. For comparison, we select the pure MADRL method (Stage I of our method) and the greedy algorithm. Clearly, the proposed method can achieve higher security performance than the pure MADRL method and greedy algorithm. Despite the introduction of multiple neural networks, the proposed method can attain a convergence speed similar to that of pure MADRL. However, due to the lack of adversarial signal design, pure MADRL security performance is relatively poor. Although the greedy algorithm has the simplest structure and process, it is difficult to optimize the network security performance with increasing training episodes. Furthermore, even though it comes at the cost of introducing a complex network structure, our proposed method can learn near-optimal policies after multiple training and deliver reliable convergence with stable gains, which proves the effectiveness of our proposed method.

\section{Conclusions and Future Work}
In this paper, a multi-agent-driven cognitive secure communication method for STNs that coordinates link scheduling and spectrum protection under real-time sensing has been presented, where the proposed method can support security requirements against new threats such as intelligent detection and interference. We have employed multi-agent coordination schedule to establish the foundation layer and determine the optimal frequency slot occupation for all UEs. Then, we have introduced a GANs-generated adversarial signal and a learning-aided power controller for protection layer. By adopting this method, the influence of non-cooperative interference can be effectively avoided, and the judgment of eavesdroppers can be confused without affecting the transmission of legitimate data. The method design and theoretical analysis in this work can provide valuable guidance on enhancing network security against evolving threats. Future work will explore hierarchical DRL and introduce friendly interference sources to improve the network security and resource utilization.

\end{document}